\documentclass{ws-mplb}
\usepackage{bm}
\bmdefine{\boldzero}{0}
\bmdefine{\bolds}{s}
\bmdefine{\boldi}{i}
\bmdefine{\boldj}{j}
\bmdefine{\boldv}{v}
\bmdefine{\boldtau}{\tau}
\bmdefine{\boldsigma}{\sigma}
\bmdefine{\boldlambda}{\lambda}
\bmdefine{\boldx}{x}
\bmdefine{\boldX}{X}
\bmdefine{\boldk}{k}
\bmdefine{\boldK}{K}
\bmdefine{\boldq}{q}
\bmdefine{\boldQ}{Q}
\bmdefine{\boldr}{r}
\bmdefine{\boldj}{j}
\begin{document}
\markboth{Naoya Arakawa}
{Many-body effects on the resistivity of a multiorbital system 
beyond Landau's Fermi-liquid theory}

%
\catchline{}{}{}{}{}
%

\title{
Many-body effects on the resistivity of a multiorbital system 
\\beyond Landau's Fermi-liquid theory
}

\author{Naoya Arakawa}

\address{Center for Emergent Matter Science (CEMS), 
RIKEN, Wako, Saitama 351-0198, Japan}

\maketitle

\begin{history}
\received{(Day Month Year)}
\revised{(Day Month Year)}
\end{history}

\begin{abstract}
I review many-body effects on the resistivity of a multiorbital system 
beyond Landau's Fermi-liquid (FL) theory. 
Landau's FL theory succeeds in describing electronic properties of 
some correlated electron systems at low temperatures. 
However, 
the behaviors deviating from the temperature dependence in the FL, 
non-FL-like behaviors, 
emerge near a magnetic quantum-critical point. 
These indicate the importance of many-body effects beyond Landau's FL theory. 
Those effects in multiorbital systems have been little understood, 
although their understanding is important to deduce 
ubiquitous properties of correlated electron systems 
and characteristic properties of multiorbital systems. 
To improve this situation, 
I formulate the resistivity of a multiorbital Hubbard model 
using the extended \'{E}liashberg theory 
and adopt this method to the inplane resistivity of 
quasi-two-dimensional paramagnetic ruthenates 
in combination with the fluctuation-exchange approximation 
including the current vertex corrections 
arising from the self-energy and Maki-Thompson term. 
The results away from and near the antiferromagnetic quantum-critical point 
reproduce 
the temperature dependence observed in Sr$_{2}$RuO$_{4}$ and Sr$_{2}$Ru$_{0.075}$Ti$_{0.025}$O$_{4}$, 
respectively. 
I highlight the importance of 
not only the momentum and the temperature dependence of the damping of a quasiparticle 
but also its orbital dependence in discussing 
the resistivity of correlated electron systems.  
\end{abstract}

\keywords{many-body effects; non-Fermi-liquid-like behaviors; 
ruthenates; $t_{2g}$ orbital; nearly magnetic metal;  
fluctuation-exchange approximation; current vertex correction.}

\section{Introduction}
Landau's Fermi-liquid (FL) theory\cite{Landau,Nozieres} can describe 
electronic properties of some correlated electron systems 
at low temperatures\cite{Leggett,Mackenzie}. 
In this theory, 
low-energy excitations are described by quasiparticles (QPs), 
i.e. electrons in a self-consistent field of surrounding electrons due to electron correlation. 
Also, 
the interactions between QPs, described by the Landau parameters, 
are independent of temperature. 
Due to these two properties, 
the temperature dependences of physical quantities in low-$T$ region 
are governed by the temperature dependence of the Fermi distribution function, 
and the corrections due to electron correlation, 
the mass enhancement and the FL correction, 
are independent of temperature. 
As a result, 
thermodynamic or magnetic or transport quantities 
show the same temperature dependences as 
those of the free electron system in low-$T$ region, 
and the difference is the coefficient 
due to the mass enhancement or the FL correction or both. 
For example, 
the specific heat is proportional to $T$, 
and the coefficient is renormalized by the mass enhancement; 
the spin susceptibility is independent of temperature, 
and the coefficient is renormalized by the mass enhancement and the FL correction.  

The original Landau phenomenological theory\cite{Landau} can be justified 
by microscopic perturbation theory with several imposed conditions\cite{Nozieres}. 

One of the basic assumptions, the one-to-one correspondence, 
becomes valid if the QP damping is much smaller than temperature considered 
(i.e., the QP lifetime, the inverse of the QP damping, is very long). 
Actually, 
the single-particle spectral function near the Fermi level 
becomes delta-function-type 
in the coherent limit [i.e., $\gamma_{\alpha}^{\ast}(\boldk)/T\rightarrow 0$ 
for all Fermi momenta], 
\begin{align}
A_{\alpha}(k)&=
-\dfrac{1}{\pi}\textrm{Im}G_{\alpha}^{(\textrm{R})}(k)
=  
\dfrac{z_{\alpha}(\boldk)}{\pi}\dfrac{\gamma_{\alpha}^{\ast}(\boldk)}
{[\epsilon-\xi_{\alpha}^{\ast}(\boldk)]^{2}+\gamma_{\alpha}^{\ast}(\boldk)^{2}}
\rightarrow   
z_{\alpha}(\boldk)\delta(\epsilon-\xi_{\alpha}^{\ast}(\boldk)).\label{eq:Akw-FL}
\end{align}
Here we consider only the coherent part, 
$G_{\alpha}^{(\textrm{R})}(k)=  
\frac{z_{\alpha}(\boldk)}{\epsilon-\xi_{\alpha}^{\ast}(\boldk)+i\gamma_{\alpha}^{\ast}(\boldk)}$, 
where the self-energy is approximated as 
$\Sigma_{\alpha}^{(\textrm{R})}(k)\approx \Sigma_{\alpha}^{(\textrm{R})}(\boldk,0)
+\epsilon\frac{\partial \Sigma_{\alpha}^{(\textrm{R})}(k)}
{\partial \epsilon}|_{\epsilon\rightarrow 0}$; 
$k$ is $k\equiv (\boldk,\epsilon)$, 
$\alpha$ is the QP band index, 
$\xi_{\alpha}^{\ast}(\boldk)$ is the QP energy, which is of the order of $T$, 
$z_{\alpha}(\boldk)=[1
-\frac{\partial \textrm{Re}\Sigma_{\alpha}^{(\textrm{R})}(\boldk,\omega)}
{\partial \omega}|_{\omega \rightarrow 0}]^{-1}$ 
is the inverse of the mass enhancement factor, 
and $\gamma_{\alpha}^{\ast}(\boldk)=
-z_{\alpha}(\boldk)\textrm{Im}\Sigma_{\alpha}^{(\textrm{R})}(\boldk,0)$ 
is the QP damping. 
Since the delta-function-type spectral function is obtained for an exact eigenstate, 
the FL becomes an approximate eigenstate for momenta near the Fermi level 
if the QP dampings at these momenta are much smaller than $T$. 
In the FL theory 
the QP damping becomes small rapidly compared with decreasing temperature 
as a result of its $T^{2}$ dependence\cite{Morel-Nozieres,AGD}. 

In addition, 
the other basic assumption about the Landau parameters becomes valid 
if the reducible four-point vertex functions and mass enhancement factor
are independent of temperature\cite{Nozieres,AGD}. 
Note, first, that the Landau parameter is proportional to 
the product of the reducible four-point vertex function 
and the square root of the four mass enhancement factors\cite{Nozieres,AGD}; 
second, that the four-point vertex function describes 
the multiple scattering of an electron and a hole\cite{Nozieres}; 
third, that 
all reducible diagrams can be split into two parts 
by removing a pair of the single-particle Green's functions, 
while all irreducible diagrams cannot be done by using that removing\cite{Nozieres}.

In contrast to successful description\cite{Mackenzie} 
in the paramagnetic (PM) phase of Sr$_{2}$RuO$_{4}$, 
Landau's FL theory fails in describing electronic properties of other PM ruthenates 
near a magnetic quantum-critical point (QCP). 
For example, 
in Sr$_{2}$Ru$_{0.075}$Ti$_{0.025}$O$_{4}$, 
located near an antiferromagnetic (AF) QCP\cite{Neutron-Ti214}, 
the spin susceptibility shows the Curie-Weiss-like temperature dependence 
and the inplane resistivity, $\rho_{ab}$, shows 
the $T$-linear dependence\cite{Ti214-nFL1,Ti214-nFL2}, 
deviating from the FL-type $T^{2}$ dependence; 
the characteristic wave vectors of this AF QCP\cite{Neutron-Ti214}, 
$\boldq\approx(\frac{2\pi}{3}, \frac{2\pi}{3})$ and its symmetrically equivalent ones, 
are the same as the wave vectors of the most strongly enhanced 
spin fluctuation in Sr$_{2}$RuO$_{4}$\cite{Neutron-x2}. 
Also, 
in Ca$_{2-x}$Sr$_{x}$RuO$_{4}$ around $x=0.5$, 
located near a ferromagnetic QCP\cite{Neutron-CSRO}, 
the Curie-Weiss-like temperature dependence of the spin susceptibility 
and $T^{1.4}$ dependence of $\rho_{ab}$ are observed\cite{CSRO-nFL1}. 

Since such non-FL-like behaviors are observed 
in other systems near a magnetic QCP\cite{Kontani-review} or a Mott transition\cite{MIT-review}, 
where many-body effects generally become very important, 
these experimental facts\cite{Ti214-nFL1,Ti214-nFL2,CSRO-nFL1} 
indicate the necessity of both 
considering how the basic assumptions of Landau's FL theory 
are violated 
and discussing many-body effects beyond Landau's FL theory 
in a more elaborate theory.

So far, 
there are two candidates for the origin of such non-FL-like behaviors. 

One is bad metal\cite{Georges-review,Kotliar-review} 
due to local correlation enhanced near a Mott transition. 
If the low-energy excitations can be satisfactorily described 
by using only low-order Taylor series of 
the self-energy in terms of frequency, 
the coherent part of the single-particle Green's function 
plays dominant roles in discussing electronic properties. 
However, 
if there are some non-negligible contributions from the higher-order Taylor series, 
the incoherent part also becomes important. 
Such case is realized in a PM metallic phase near the Mott transition 
due to the formation of the upper and lower Hubbard peaks, 
arising from enhanced local correlation\cite{Hubbard,DMFT-1st}. 
Since perturbation theory can treat the coherent part appropriately 
and its treatment of the incoherent part is insufficient\cite{Ikeda-4th}, 
perturbation theory is unsuitable in the latter case. 
Instead, 
the latter case can be well described 
in dynamical-mean-field theory (DMFT) 
since the DMFT can take account of the frequency dependence of the self-energy 
nonperturbatively\cite{Georges-review,Kotliar-review}. 
Actually, 
several non-FL-like behaviors are obtained in the DMFT 
near the Mott transition 
as a result of the $T$-linear unrenormalized QP damping and 
the temperature-dependent mass enhancement factor\cite{Prus,DMFT-trans-QP}. 
[In this paper, 
I define the unrenormalized QP damping as 
the QP damping divided by the inverse of the mass enhancement factor.] 
These results indicate 
the importance of nonperturbative effects of local correlation 
near the Mott transition.
Note that 
Hund's metal\cite{Haule-FePn,Georges-Hund} is classified as the bad metal 
since the non-FL-like behaviors in the Hund's metal 
arise from local spin fluctuation enhanced near a Mott transition, 
although there is a crucial difference between 
the Hund's rule coupling dependence of 
the critical value of the intraorbital Coulomb interaction for the Mott transition 
at half-filling and non-half-filling\cite{Georges-Hund}.  

The other is nearly magnetic metal\cite{NearlyFM1,NearlyFM2,NearlyAF,NearlyFM-trans} 
due to spatial correlation enhanced near a magnetic QCP. 
If a system is located far away from a magnetic QCP, 
all scattering processes are independent of temperature. 
However, 
if the system approaches the QCP, 
several characteristic scattering processes of the QCP show 
the strong temperature-dependent enhancement\cite{Kon-CVC}. 
For example, 
in case near a stripe-type AF QCP 
the scattering processes mediated by AF spin fluctuations with 
$\boldq=(\pi,\pi)$ and its symmetrically equivalent ones 
are strongly enhanced as decreasing temperature\cite{Kon-CVC}. 
Such strong temperature-dependent enhancement leads to 
the strong temperature dependence of the reducible four-point vertex function 
whose momentum is characterized by the QCP. 
Thus, the basic assumption of Landau's FL theory about the Landau parameters 
is violated only for the characteristic momenta of spin fluctuation 
enhanced near the QCP\cite{Kon-CVC}. 
Also, the basic assumption about the QP damping is violated 
for the characteristic momenta due to the formation of hot spot, 
arising from enhanced spin fluctuation\cite{Kon-CVC}; 
at the hot spot, 
the QP damping does not become much smaller than temperature considered. 
For example, 
in case near the stripe-type AF QCP, 
the QP dampings at the momenta connected by the nesting vector $\boldq=(\pi,\pi)$ 
and its symmetrically equivalent ones 
are more strongly enhanced than those at the other momenta 
due to the enhancement of the corresponding AF spin fluctuations\cite{Kon-CVC}. 
These two violations suggest the necessity of  
discussing electronic properties near a magnetic QCP 
in the microscopic perturbation theory beyond Landau's FL theory. 
Actually, 
fluctuation-exchange (FLEX) approximation reproduces 
several non-FL-like behaviors 
due to the hot-spot structure of the QP damping or 
the Curie-Weiss-like temperature dependence of spin fluctuations or both\cite{Kon-CVC,NA}. 
As explained in Sect. 2.3, 
this approximation can take account of spatial correlation 
beyond a mean-field theory 
and describe electronic properties of a metallic phase at low temperatures 
for moderately strong electron correlation satisfactorily\cite{FLEX1,FLEX2,FLEX3}. 
In contrast to case near the Mott transition, 
the DMFT is inappropriate to describe electronic properties near a magnetic QCP 
since the DMFT neglects spatial correlation completely\cite{Metzner}. 
These results indicate powerfulness of the microscopic perturbation theory 
and the importance of temperature-dependent spatial correlation near a magnetic QCP. 

With the backgrounds explained above, 
I studied electronic structure and magnetic and transport properties\cite{NA} 
of ruthenates near and away from the AF QCP 
in the FLEX approximation with 
current vertex corrections (CVCs) 
arising from the self-energy and 
Maki-Thompson (MT)\cite{MT1,MT2} term 
for the $t_{2g}$-orbital Hubbard model on a square lattice 
and succeeded in reproducing 
several experimental results of 
Sr$_{2}$Ru$_{0.075}$Ti$_{0.025}$O$_{4}$\cite{Neutron-Ti214,Ti214-nFL1,Ti214-nFL2} 
and Sr$_{2}$RuO$_{4}$\cite{Neutron-x2,dHvA-x2,resistivity-x2,Hall-x2}. 
Thus, 
the non-FL-like behaviors\cite{Ti214-nFL1,Ti214-nFL2} 
in Sr$_{2}$Ru$_{0.075}$Ti$_{0.025}$O$_{4}$ 
can be understood as the nearly magnetic meal near the AF QCP. 
Moreover, 
since the results away from the AF QCP about 
the orbital dependence of the mass enhancement 
are in better agreement with the experiment\cite{dHvA-x2} in Sr$_{2}$RuO$_{4}$ 
than those of the DMFT\cite{Haule-DMFT}, 
electronic properties of ruthenates at low temperatures 
except a few cases\cite{Ca214,Ca327} near the Mott transition 
may be better described in the microscopic perturbation theory than in the DMFT. 

In this paper, 
I review part of the above previous study\cite{NA} and show some new results. 
In Sect. 2, 
I explain the microscopic theory\cite{NA} used for analyzing $\rho_{ab}$ of 
some quasi-two-dimensional (quasi-$2$D) PM ruthenates.  
In Sect. 3, 
I show the results about many-body effects on $\rho_{ab}$ 
of the ruthenates near and away from the AF QCP 
in the FLEX approximation 
with the CVCs arising from the self-energy and MT term 
and compare these results with the results obtained in other cases 
where the CVCs are more simplified. 
In Sect. 4, 
I summarize the results and draw some conclusions. 

Due to the limit of space, 
I do not consider the Aslamasov-Larkin (AL) CVC\cite{AL}, 
which is the other CVC in the FLEX approximation; 
its detailed derivation and effects are going to be discussed elsewhere\cite{NA-full}.

\section{Method}
In this section, 
I explain an effective model of some quasi-$2$D ruthenates, 
briefly review the formal derivation of the resistivity of a multiorbital Hubbard model 
in a PM metallic phase, 
and formulate the microscopic perturbation theory used to calculate the resistivity. 
The more detailed explanations 
about those derivation and formulation are going to be given elsewhere\cite{NA-full}. 

In the following, 
I use the unit $\hbar=c=e=\mu_{\textrm{B}}=k_{\textrm{B}}=1$, 
set the coordinates $x$, $y$, and $z$ 
in the directions of the Ru$-$O bonds of a RuO$_{6}$ octahedral, 
and label the $d_{xz}$, $d_{yz}$, and $d_{xy}$ orbitals as $1$, $2$, and $3$, respectively. 

\subsection{Effective model for some quasi-$2$D ruthenates}
I explain an effective model 
to describe electronic properties of some quasi-$2$D ruthenates 
without the rotation or the tilting of RuO$_{6}$ octahedra\cite{Rot-Tilt}. 

Before introducing the effective model, 
I briefly explain several basic electronic properties. 
Some ruthenates whose crystal structures are 
$214$-type such as Sr$_{2}$RuO$_{4}$\cite{Mackenzie} or 
$327$-type such as Sr$_{3}$Ru$_{2}$O$_{7}$\cite{Ru327-1} 
are categorized into quasi-$2$D $t_{2g}$-orbital systems. 
For simplicity, we focus on Sr$_{2}$RuO$_{4}$; 
the following properties remain qualitatively the same 
in other metallic ruthenates\cite{CSRO-nFL1,Ru327-1}.  
First, 
the inplane resistivity 
is about $10^{-3}$ times as small as the out-of-plane resistivity 
at low temperatures\cite{Maeno-triplet}, 
and  
the almost cylindrical Fermi surface (FS) 
is observed in the de Haas-van Alphen measurement\cite{dHvA-x2}. 
These indicate quasi-$2$D electronic conduction. 
Moreover, 
according to several density-functional calculations\cite{Oguchi,Mazin} 
in local-density approximation (LDA), 
conducting bands near the Fermi level 
are formed by the antibonding bands of the Ru $t_{2g}$ and the O $2p$ orbitals, 
and 
the $t_{2g}$ orbitals mainly contribute to 
the density-of-states (DOS) near the Fermi level. 
Since the topology of the FS obtained in the LDA 
qualitatively agrees with experiments\cite{dHvA-x2,ARPES-x2}, 
the $t_{2g}$ orbitals play dominant roles in discussing electronic properties 
at low temperatures. 

With the above background, 
I assume that 
the electronic structure obtained in the LDA\cite{Oguchi,Mazin} for Sr$_{2}$RuO$_{4}$ 
is a good starting point to 
consider many-body effects beyond a mean-field approximation, 
and I use a $t_{2g}$-orbital Hubbard model on a square lattice as the effective model. 
Thus, the Hamiltonian is $\hat{H}=\hat{H}_{0}+\hat{H}_{\textrm{int}}$ with 
\begin{align}
\hat{H}_{0}
&=
\textstyle\sum\limits_{\boldk}
\textstyle\sum\limits_{a,b=1}^{3}
\textstyle\sum\limits_{s=\uparrow,\downarrow}
\epsilon_{ab}(\boldk)
\hat{c}^{\dagger}_{\boldk a s} 
\hat{c}_{\boldk b s},\label{eq:H0}
\end{align}
and
\begin{align}
\hat{H}_{\textrm{int}}
&=\dfrac{1}{4}
\textstyle\sum\limits_{\boldj}
\textstyle\sum\limits_{a,b,c,d=1}^{3}
\textstyle\sum\limits_{s_{1},s_{2},s_{3},s_{4}=\uparrow,\downarrow}
U_{abcd}^{s_{1}s_{2}s_{3}s_{4}}
\hat{c}^{\dagger}_{\boldj a s_{1}}\hat{c}^{\dagger}_{\boldj d s_{4}}
\hat{c}_{\boldj c s_{3}}\hat{c}_{\boldj b s_{2}}\notag\\
&=
 U 
\textstyle\sum\limits_{\boldj}
\textstyle\sum\limits_{a=1}^{3}
\hat{n}_{\boldj a \uparrow} \hat{n}_{\boldj a \downarrow}
+ U^{\prime}  
\textstyle\sum\limits_{\boldj}
\textstyle\sum\limits_{a=1}^{3}
\textstyle\sum\limits_{b<a}
\hat{n}_{\boldj a} \hat{n}_{\boldj b}\notag\\
&- 
J_{\textrm{H}} 
\textstyle\sum\limits_{\boldj}
\textstyle\sum\limits_{a=1}^{3}
\textstyle\sum\limits_{b<a}
( 
2 \hat{\bolds}_{\boldj a} \cdot 
\hat{\bolds}_{\boldj b} 
+ 
\frac{1}{2} \hat{n}_{\boldj a} \hat{n}_{\boldj b} 
)+
J^{\prime} 
\textstyle\sum\limits_{\boldj}
\textstyle\sum\limits_{a=1}^{3}
\textstyle\sum\limits_{b\neq a}
\hat{c}_{\boldj a \uparrow}^{\dagger} 
\hat{c}_{\boldj a \downarrow}^{\dagger} 
\hat{c}_{\boldj b \downarrow} 
\hat{c}_{\boldj b \uparrow},\label{eq:Hint}
\end{align}
where $\epsilon_{ab}(\boldk)$ is the energy dispersions, 
measuring from the chemical potential, $\mu$, 
$U$, $U^{\prime}$, $J_{\textrm{H}}$, and $J^{\prime}$ are 
intraorbital Coulomb interaction, 
interorbital Coulomb interaction, 
Hund's rule coupling, and pair hopping term, 
$\hat{n}_{\boldj a}$ is 
$\hat{n}_{\boldj a}=\sum_{s}\hat{n}_{\boldj a s}=
\sum_{s}\hat{c}^{\dagger}_{\boldj a s}\hat{c}_{\boldj a s}$ 
and $\hat{\bolds}_{\boldj a}$ is  $\hat{\bolds}_{\boldj a}=\frac{1}{2}
\sum_{s,s^{\prime}}\hat{c}^{\dagger}_{\boldj a s} 
\boldsigma_{s s^{\prime}} \hat{c}_{\boldj a s^{\prime}}$ 
with the Pauli matrices $\boldsigma_{s s^{\prime}}$. 
Since I do not consider the effects\cite{NA-GA} of the rotation and the tilting, 
we focus on electronic properties of Sr$_{2}$RuO$_{4}$\cite{Mackenzie} 
and some doped Sr$_{2}$RuO$_{4}$\cite{Ti214-nFL1} without these distortions. 
Note that the rotation is present in Ca$_{2-x}$Sr$_{x}$RuO$_{4}$ around $x=0.5$\cite{Rot-Tilt}. 

\begin{figure}[tb]
\begin{center}
\includegraphics[width=120mm]{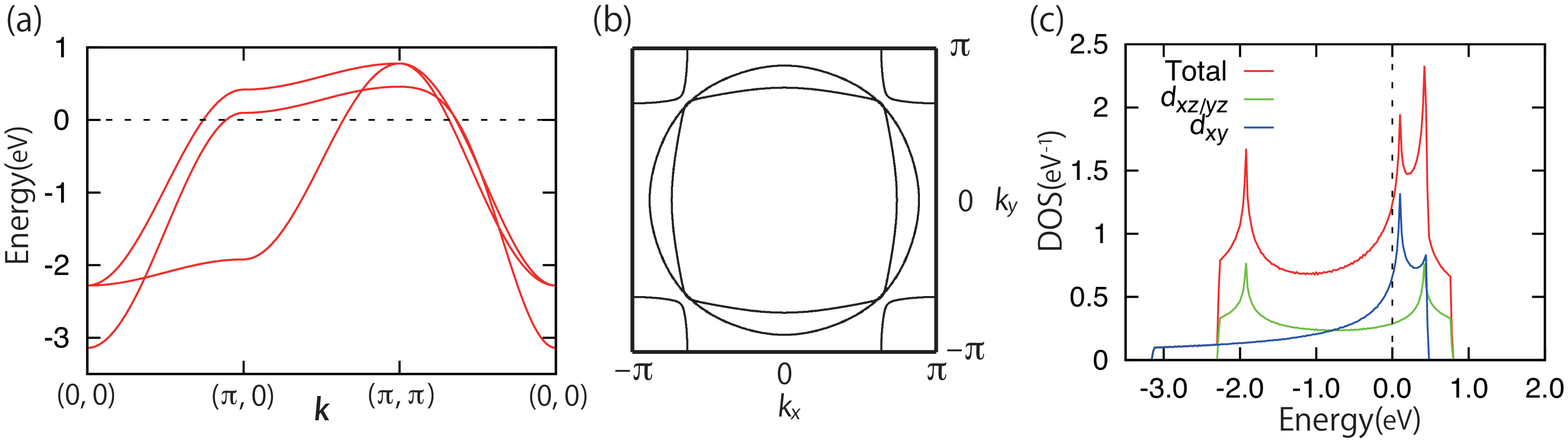}
\end{center}
\vspace{-20pt}
\caption{(a) Band structure, (b) FS, and (c) DOS for $\hat{H}_{0}$ whose parameters are chosen 
so as to reproduce the results in the LDA$^{47,48}$ for Sr$_{2}$RuO$_{4}$.}
\label{fig:Fig1}
\end{figure}

By considering some symmetrically possible hopping processes 
and the difference between the crystalline-electric-field energies of 
the $d_{xy}$ and $d_{xz/yz}$ orbitals, $\Delta_{t_{2g}}$, 
we can construct the tight-binding model whose $\epsilon_{ab}(\boldk)$ is given by 
\begin{align}
\epsilon_{11}(\boldk)=&
-\frac{\Delta_{t_{2g}}}{3}-2 t_{1} \cos k_{x}-2 t_{2} \cos k_{y}-\mu,\\ 
\epsilon_{12}(\boldk)=&\ \epsilon_{21}(\boldk)=\
 4 t^{\prime} \sin k_{x} \sin k_{y},\\ 
\epsilon_{22}(\boldk)=&
-\frac{\Delta_{t_{2g}}}{3}-2 t_{2} \cos k_{x}-2 t_{1} \cos k_{y}-\mu,\\ 
\epsilon_{33}(\boldk)=&\
\frac{2\Delta_{t_{2g}}}{3}-2t_{3}(\cos k_{x}+\cos k_{y})-4t_{4}\cos k_{x} \cos k_{y}-\mu,
\end{align}
and otherwise $\epsilon_{ab}(\boldk)=0$. 
$\mu$ is chosen so that 
the electron number per a site, $n_{\textrm{e}}$, is fixed. 
For the actual calculations, 
$\mu$ is determined by the bisection method using 
\begin{align}
n_{\textrm{e}}= 
\dfrac{2}{N}
\textstyle\sum\limits_{\boldk}
\textstyle\sum\limits_{\alpha}
f(\epsilon_{\alpha}(\boldk))
+
\dfrac{2T}{N}
\textstyle\sum\limits_{\boldk}
\textstyle\sum\limits_{m}
\textstyle\sum\limits_{a=1}^{3}
\Bigl[
G_{aa}(\boldk,i\omega_{m})-G_{aa}^{0}(\boldk,i\omega_{m})
\Bigr],\label{eq:mu-int}
\end{align}
where the second term becomes zero without the interaction terms. 
Here 
$\epsilon_{\alpha}(\boldk)$ is 
\begin{align}
\epsilon_{\alpha}(\boldk)
=
\textstyle\sum\limits_{a,b=1}^{3}
(U_{\boldk}^{0 \dagger})_{\alpha a}
\epsilon_{ab}(\boldk)
(U_{\boldk}^{0})_{b\alpha},
\end{align}
$f(\epsilon)$ is the Fermi distribution function, 
$G_{ab}^{0}(\boldk,i\omega_{m})$ is the noninteracting single-particle Green's function 
with fermionic Matsubara frequency, $\omega_{m}=\pi T(2m+1)$, 
\begin{align}
G_{ab}^{0}(\boldk,i\omega_{m})=
\textstyle\sum\limits_{\alpha}
(U_{\boldk}^{0})_{a\alpha}\dfrac{1}{i\omega_{m}-\epsilon_{\alpha}(\boldk)}
(U_{\boldk}^{0 \dagger})_{\alpha b},\label{eq:G0}
\end{align} 
and $G_{ab}(\boldk,i\omega_{m})$ is the single-particle Green's function 
whose determination is explained in Sect. 2.3. 
In Eq. (\ref{eq:mu-int}), 
we put the chemical potentials in $f(\epsilon_{\alpha}(\boldk))$, $G_{aa}(\boldk,i\omega_{m})$, 
and $G_{aa}^{0}(\boldk,i\omega_{m})$ the same 
to reduce the numerical error arising from the cut-off frequency. 
To reproduce the electronic structure obtained in the LDA\cite{Oguchi,Mazin}, 
I set 
$(t_{1},t_{2},t_{3},t_{4},t^{\prime},\Delta_{t_{2g}})=
(0.675, 0.09, 0.45, 0.18, 0.03, 0.13)$ (eV) and $n_{\textrm{e}}=4$. 
Actually, we see from Fig. \ref{fig:Fig1} 
that 
the total bandwidth, 
the topology of each FS sheet, 
and the location of the van Hove singularity of the $d_{xy}$ orbital 
agree with the LDA results\cite{Oguchi,Mazin}: 
the total bandwidth is about $4$ eV; 
the FS consists of 
the quasi-$1$D hole-like $\alpha$ and electron-like $\beta$ sheets 
of the $d_{xz/yz}$ orbital and 
the quasi-$2$D electron-like $\gamma$ sheet of the $d_{xy}$ orbital, 
and the $\gamma$ sheet is located nearer the inner sheet in $k_{x}=k_{y}$ or $k_{x}=-k_{y}$ line 
than in the experiment\cite{ARPES-x2}; 
the van Hove singularity is located above the Fermi level. 
Also, 
the occupation numbers of the $d_{xz/yz}$ and the $d_{xy}$ orbital, 
being $n_{xz/yz}=1.38$ and $n_{xy}=1.25$, 
are consistent with the LDA values\cite{Oguchi,Mazin}. 

Then, 
I set $J^{\prime}=J_{\textrm{H}}$, $U^{\prime}=U-2J_{\textrm{H}}$, 
and $J_{\textrm{H}}=\frac{U}{6}$, use $U$ as a parameter,  
and treat the effects of interactions 
in the FLEX approximation with 
the CVCs arising from the self-energy and MT term; 
its detail is explained in Sect. 2.3. 

Finally, 
we remark on suitability neglecting the spin-orbit coupling of Ru ions 
for discussing many-body effects on the resistivity. 
A density-functional calculation\cite{Oguchi-LS} for Sr$_{2}$RuO$_{4}$ 
within local-spin-density approximation shows that 
the coupling constant is $0.167$ eV, 
and that the main effect on the electronic structure is the weak mixing 
between the bands of the $d_{xz/yz}$ and the $d_{xy}$ orbital 
around $\boldk=(\frac{2\pi}{3},\frac{2\pi}{3})$ and its symmetrically equivalent ones. 
This effect will not qualitatively change the results shown in Sect. 3 
since this coupling constant is smaller than the main terms in $\hat{H}$ 
and that weak mixing will lead to small changes of the momentum dependence and value 
of the spin susceptibility from those without the spin-orbit coupling. 
Thus, I believe that 
neglecting the spin-orbit coupling is suitable 
for qualitative discussions about many-body effects on the resistivity. 

\subsection{Extended  \'{E}liashberg theory for the resistivity of 
  a multiorbital Hubbard model in a PM metallic phase}
I briefly review the formal derivation\cite{NA,NA-full} of the resistivity 
of a multiorbital Hubbard model in a PM metallic phase. 
We first derive an exact expression of the longitudinal conductivity, 
the inverse of the resistivity, 
in the presence of electron correlation 
within the linear-response theory\cite{Kubo-formula}. 
Then, we rewrite this exact expression in terms of the four-point vertex function 
by using the three-point vector vertex function. 
Due to difficulty solving the exact expression, 
we derive an approximate expression 
by using the most-divergent-term approximation introduced by \'{E}liashberg\cite{Eliashberg}.

To discuss the resistivity within the linear-response theory, 
we use the Kubo formula\cite{Kubo-formula} for the longitudinal conductivity, 
$\sigma_{\nu \nu}$ ($\nu=x,y$), in $\omega$-limit and $\omega\tau_{\textrm{trans}} \ll 1$ 
with $\tau_{\textrm{trans}}$ being the transport relaxation time\cite{Eliashberg}, 
which is of the order of magnitude of the QP lifetime. 
Namely, $\sigma_{\nu\nu}$ is given by 
\begin{align}
\sigma_{\nu\nu}=
2
\textstyle\lim\limits_{\omega\rightarrow 0}
\textstyle\lim\limits_{\boldq\rightarrow \boldzero}
\dfrac{\tilde{K}_{\nu \nu}^{(\textrm{R})}(\boldq,\omega)
-\tilde{K}_{\nu \nu}^{(\textrm{R})}(\boldq,0)}{i\omega}
=
2
\textstyle\lim\limits_{\omega\rightarrow 0}
\dfrac{\tilde{K}_{\nu \nu}^{(\textrm{R})}(\boldzero,\omega)
-\tilde{K}_{\nu \nu}^{(\textrm{R})}(\boldzero,0)}{i\omega},\label{eq:sigmaxx}
\end{align}
where $\tilde{K}_{\nu \nu}^{(\textrm{R})}(\boldzero,\omega)$ is obtained 
by the analytic continuation of $\tilde{K}_{\nu \nu}(i\Omega_{n})$, 
\begin{align}
\tilde{K}_{\nu \nu}(i\Omega_{n})
=&\
\textstyle\lim\limits_{\boldq\rightarrow \boldzero}
\dfrac{1}{N}
\int^{T^{-1}}_{0}d\tau e^{i\Omega_{n}\tau}
\langle \textrm{T}_{\tau}  
\hat{J}_{\boldq \nu}(\tau)
\hat{J}_{-\boldq \nu}(0)\rangle\notag\\
=&\ 
\frac{1}{N}
\textstyle\sum\limits_{\boldk,\boldk^{\prime}}
\textstyle\sum\limits_{\{a\}}
\int^{T^{-1}}_{0}d\tau e^{i\Omega_{n}\tau}
(v_{\boldk \nu})_{ba}
(v_{\boldk^{\prime} \nu})_{cd}
\langle \textrm{T}_{\tau}  
\hat{c}_{\boldk b}^{\dagger}(\tau)
\hat{c}_{\boldk a}(\tau)
\hat{c}_{\boldk^{\prime} c}^{\dagger}
\hat{c}_{\boldk^{\prime} d} \rangle\notag\\
=&\ 
\frac{1}{N}
\textstyle\sum\limits_{\boldk,\boldk^{\prime}}
\textstyle\sum\limits_{\{a\}}
(v_{\boldk \nu})_{ba}
(v_{\boldk^{\prime} \nu})_{cd}
K_{abcd}(\boldk,\boldk^{\prime};i\Omega_{n}),\label{eq:Ktild}
\end{align}
with bosonic Matsubara frequency, $\Omega_{n}=2\pi T n$. 
Here 
$\sum_{\{a\}}$ is $\sum_{\{a\}}\equiv \sum_{a,b,c,d}$, 
$(v_{\boldk \nu})_{ab}$ is the group velocity, 
$(v_{\boldk \nu})_{ab}=\frac{\partial \epsilon_{ab}(\boldk)}{\partial k_{\nu}}$, 
and $K_{abcd}(\boldk,\boldk^{\prime};i\Omega_{n})$ is 
\begin{align}
K_{abcd}(\boldk,\boldk^{\prime};i\Omega_{n})
=& 
-\delta_{\boldk,\boldk^{\prime}}
T\textstyle\sum\limits_{m}
G_{ac}(\boldk,i\omega_{m+n})
G_{db}(\boldk,i\omega_{m})\notag\\
&-T^{2}\textstyle\sum\limits_{m,m^{\prime}}
\textstyle\sum\limits_{\{A\}}
G_{aA}(\boldk,i\omega_{m+n})
G_{dD}(\boldk^{\prime},i\omega_{m^{\prime}})
G_{Bb}(\boldk,i\omega_{m})\notag\\
&\ \times 
G_{Cc}(\boldk^{\prime},i\omega_{m^{\prime}+n})
\Gamma_{\{A\}}(\boldk,i\omega_{m},\boldk^{\prime},i\omega_{m^{\prime}};i\Omega_{n}),\label{eq:2G-Matsu}
\end{align}
where $\Gamma_{\{A\}}(\boldk,i\omega_{m},\boldk^{\prime},i\omega_{m^{\prime}};i\Omega_{n})\equiv 
\Gamma_{ABCD}(\boldk,i\omega_{m},\boldk^{\prime},i\omega_{m^{\prime}};i\Omega_{n})$ 
is the reducible four-point vertex function. 
Thus, the analytic continuation of $K_{abcd}(\boldk,\boldk^{\prime};i\Omega_{n})$ 
is necessary to calculate $\sigma_{\nu\nu}$. 

Before the analytic continuation of $K_{abcd}(\boldk,\boldk^{\prime};i\Omega_{n})$, 
I remark on the important physical meanings of 
$\lim_{\omega\rightarrow 0}\lim_{\boldq\rightarrow \boldzero}$ and $\omega\tau_{\textrm{trans}} \ll 1$. 
The order of $\lim_{\omega\rightarrow 0}$ and $\lim_{\boldq\rightarrow \boldzero}$ 
is very important in discussing transport properties 
since the observable currents can be obtained 
by the dynamical and uniform field (i.e., $\lim_{\omega\rightarrow 0}\lim_{\boldq\rightarrow \boldzero}$) 
but the static and non-uniform field (i.e., $\lim_{\boldq\rightarrow \boldzero}\lim_{\omega\rightarrow 0}$) 
does not cause any observable currents 
due to the screening induced by the modulations of the charge distribution\cite{Yamada-text}. 
Also, 
the value of $\omega\tau_{\textrm{trans}}$ is very important 
since the adiabatic condition $\omega\tau_{\textrm{trans}} \ll 1$ means 
the realization of local equilibrium 
due to the rapid relaxation compared with $\omega^{-1}$, 
a typical time scale of the field; 
as a result of that relaxation, 
the electronic transports are governed mainly by the QPs near the Fermi level.  
For example, 
such importance of the inequality of $\omega \tau_{\textrm{trans}}$ 
is seen from the difference between the zero and the first sound\cite{Nozieres-Pines}. 

\begin{figure}[tb]
\begin{center}
\hspace{36pt}
\includegraphics[width=60mm]{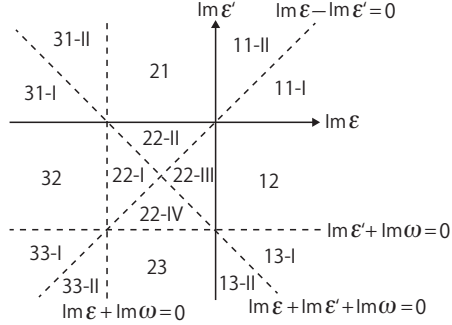}
\end{center}
\vspace{-16pt}
\caption{The connection between 
the additional subscripts of the four-point vertex function 
and the relations among $(\epsilon,\epsilon^{\prime},\omega)$ of that function.}
\label{fig:Fig2}
\end{figure}

Replacing $T\sum_{m}$ and $T^{2}\sum_{m,m^{\prime}}$ in Eq. (\ref{eq:2G-Matsu}) 
by the corresponding contour integrals\cite{AGD,Eliashberg} 
and doing several straightforward calculations\cite{NA-full} 
with attention to the analytic properties\cite{Eliashberg} of 
the single-particle Green's function and four-point vertex function, 
we can carry out the analytic continuation of $K_{abcd}(\boldk,\boldk^{\prime};i\Omega_{n})$. 
As a result, 
$\tilde{K}_{\nu \nu}^{(\textrm{R})}(\boldzero,\omega)$ is given by 
\begin{align}
\tilde{K}_{\nu \nu}^{(\textrm{R})}(\boldzero,\omega)
=&-\frac{1}{N}
\textstyle\sum\limits_{\boldk,\boldk^{\prime}}
\textstyle\sum\limits_{\{a\}}
(v_{\boldk \nu})_{ba}
(v_{\boldk^{\prime} \nu})_{cd}\notag\\
&\times 
\int^{\infty}_{-\infty}\dfrac{d\epsilon}{4\pi i}
\Bigl[
\tanh \dfrac{\epsilon}{2T}
K_{1;abcd}^{(\textrm{R})}(\boldk,\boldk^{\prime}; \epsilon; \omega)\notag\\
&\ \ \ \ \ \ \ \ \ \ \ \ \ \ \ 
+\Bigl(\tanh \dfrac{\epsilon+\omega}{2T}
-\tanh \dfrac{\epsilon}{2T}\Bigr)
K_{2;abcd}^{(\textrm{R})}(\boldk,\boldk^{\prime}; \epsilon; \omega)\notag\\
&\ \ \ \ \ \ \ \ \ \ \ \ \ \ \  
-\tanh \dfrac{\epsilon+\omega}{2T}
K_{3;abcd}^{(\textrm{R})}(\boldk,\boldk^{\prime}; \epsilon; \omega)
\Bigr],\label{eq:sum-analytic}
\end{align}
where $K_{l;abcd}^{(\textrm{R})}(\boldk,\boldk^{\prime}; \epsilon; \omega)$ is 
\begin{align}
K_{l;abcd}^{(\textrm{R})}(\boldk,\boldk^{\prime}; \epsilon; \omega)
=&\
g_{l;acdb}(k;\omega)\delta_{\boldk,\boldk^{\prime}}+
\int^{\infty}_{-\infty}\frac{d\epsilon^{\prime}}{4\pi i}
\textstyle\sum\limits_{\{A\}}
\textstyle\sum\limits_{l^{\prime}=1}^{3}
g_{l;aABb}(k;\omega)\notag\\
&\ \ \ \ \ \ \ \ \ \ \ \ \ \ \ \ \ \ \ \ \ \ \ \ \times 
\mathcal{J}_{ll^{\prime};\{A\}}(k,k^{\prime};\omega)
g_{l^{\prime};CcdD}(k^{\prime};\omega),\label{eq:Kpart-analytic}
\end{align}
with $g_{l;acdb}(k;\omega)$ being 
\begin{align}
g_{1;acdb}(k;\omega)=&\
G_{ac}^{(\textrm{R})}(\boldk,\epsilon+\omega)
G_{db}^{(\textrm{R})}(\boldk,\epsilon),\label{eq:g1}\\
g_{2;acdb}(k;\omega)=&\
G_{ac}^{(\textrm{R})}(\boldk,\epsilon+\omega)
G_{db}^{(\textrm{A})}(\boldk,\epsilon),\label{eq:g2}
\end{align}
and
\begin{align}
g_{3;acdb}(k;\omega)=&\
G_{ac}^{(\textrm{A})}(\boldk,\epsilon+\omega)
G_{db}^{(\textrm{A})}(\boldk,\epsilon),\label{eq:g3}
\end{align}
and $\mathcal{J}_{ll^{\prime};\{A\}}(k,k^{\prime};\omega)$ being 
\begin{align}
\mathcal{J}_{11;\{A\}}(k,k^{\prime};\omega)
=&\
\tanh \frac{\epsilon^{\prime}}{2T} 
\Gamma_{11\textrm{-I};\{A\}}(k,k^{\prime};\omega)\notag\\
&+\coth \frac{\epsilon^{\prime}-\epsilon}{2T} 
\Bigl[\Gamma_{11\textrm{-II};\{A\}}(k,k^{\prime};\omega)-
\Gamma_{11\textrm{-I};\{A\}}(k,k^{\prime};\omega)\Bigr],\label{eq:4VC-11}\\
\mathcal{J}_{12;\{A\}}(k,k^{\prime};\omega)
=&\
\Bigl(\tanh \frac{\epsilon^{\prime}+\omega}{2T} 
-\tanh \frac{\epsilon^{\prime}}{2T}\Bigr)
\Gamma_{12;\{A\}}(k,k^{\prime};\omega),\label{eq:4VC-12}\\
\mathcal{J}_{13;\{A\}}(k,k^{\prime};\omega)
=&
-\tanh \frac{\epsilon^{\prime}+\omega}{2T} 
\Gamma_{13\textrm{-I};\{A\}}(k,k^{\prime};\omega)\notag\\
&-\coth \frac{\epsilon+\epsilon^{\prime}+\omega}{2T} 
\Bigl[\Gamma_{13\textrm{-II};\{A\}}(k,k^{\prime};\omega)-
\Gamma_{13\textrm{-I};\{A\}}(k,k^{\prime};\omega)\Bigr],\label{eq:4VC-13}\\
\mathcal{J}_{21;\{A\}}(k,k^{\prime};\omega)
=&\
\tanh \frac{\epsilon^{\prime}}{2T} 
\Gamma_{21;\{A\}}(k,k^{\prime};\omega),\label{eq:4VC-21}\\
\mathcal{J}_{22;\{A\}}(k,k^{\prime};\omega)
=&\
\Bigl(\coth \frac{\epsilon^{\prime}-\epsilon}{2T}
-\tanh \frac{\epsilon^{\prime}}{2T}\Bigr)
\Gamma_{22\textrm{-II};\{A\}}(k,k^{\prime};\omega)\notag\\
&+
\Bigl(\coth \frac{\epsilon^{\prime}+\epsilon+\omega}{2T}
-\coth \frac{\epsilon^{\prime}-\epsilon}{2T}\Bigr)
\Gamma_{22\textrm{-III};\{A\}}(k,k^{\prime};\omega)\notag\\
&+
\Bigl(\tanh \frac{\epsilon^{\prime}+\omega}{2T}
-\coth \frac{\epsilon^{\prime}+\epsilon+\omega}{2T}\Bigr)
\Gamma_{22\textrm{-IV};\{A\}}(k,k^{\prime};\omega),\label{eq:4VC-22}\\
\mathcal{J}_{23;\{A\}}(k,k^{\prime};\omega)
=&
-\tanh \frac{\epsilon^{\prime}+\omega}{2T} 
\Gamma_{23;\{A\}}(k,k^{\prime};\omega),\label{eq:4VC-23}
\end{align}
\begin{align}
\mathcal{J}_{31;\{A\}}(k,k^{\prime};\omega)
=&\
\tanh \frac{\epsilon^{\prime}}{2T} 
\Gamma_{31\textrm{-I};\{A\}}(k,k^{\prime};\omega)\notag\\
&+\coth \frac{\epsilon+\epsilon^{\prime}+\omega}{2T} 
\Bigl[\Gamma_{31\textrm{-II};\{A\}}(k,k^{\prime};\omega)-
\Gamma_{31\textrm{-I};\{A\}}(k,k^{\prime};\omega)\Bigr],\label{eq:4VC-31}\\
\mathcal{J}_{32;\{A\}}(k,k^{\prime};\omega)
=&\
\Bigl(\tanh \frac{\epsilon^{\prime}+\omega}{2T} 
-\tanh \frac{\epsilon^{\prime}}{2T}\Bigr)
\Gamma_{32;\{A\}}(k,k^{\prime};\omega),\label{eq:4VC-32}
\end{align}
and 
\begin{align}
\hspace{-18pt}
\mathcal{J}_{33;\{A\}}(k,k^{\prime};\omega)
=&
-\tanh \frac{\epsilon^{\prime}+\omega}{2T} 
\Gamma_{33\textrm{-I};\{A\}}(k,k^{\prime};\omega)\notag\\
&-\coth \frac{\epsilon^{\prime}-\epsilon}{2T} 
\Bigl[\Gamma_{33\textrm{-II};\{A\}}(k,k^{\prime};\omega)-
\Gamma_{33\textrm{-I};\{A\}}(k,k^{\prime};\omega)\Bigr].\label{eq:4VC-33}
\end{align}
In Eq. (\ref{eq:Kpart-analytic}), 
I have not explicitly written 
whether the frequency integral is the principal integral or not; 
the integrals containing hyperbolic cotangent are the principal ones. 
Also, 
in Eqs. (\ref{eq:4VC-11}){--}(\ref{eq:4VC-33}) 
the additional subscript of the four-point vertex function such as $12$  
represents the relations among its three frequency variables, as shown in Fig. \ref{fig:Fig2}. 
Since $\mathcal{J}_{ll^{\prime};\{a\}}(k,k^{\prime};\omega)$ is determined by 
the Bethe-Salpeter equation, 
\begin{align}
\mathcal{J}_{ll^{\prime};\{a\}}(k,k^{\prime};\omega)
=&\ \mathcal{J}^{(1)}_{ll^{\prime};\{a\}}(k,k^{\prime};\omega) 
+\textstyle\sum\limits_{l^{\prime\prime}= 1}^{3}
\dfrac{1}{N}
\textstyle\sum\limits_{\boldk^{\prime\prime}}
\textstyle\sum\limits_{\{A\}}
\int^{\infty}_{-\infty}\frac{d\epsilon^{\prime\prime}}{4\pi i}
\mathcal{J}_{ll^{\prime\prime};abCD}(k,k^{\prime\prime};\omega)\notag\\
&\ \ \ \ \ \ \ \ \ \ \ \ \ \ \ \ \ \ \ \ \ \ \ \ \times
g_{l^{\prime\prime};CABD}(k^{\prime\prime};\omega)
\mathcal{J}^{(1)}_{l^{\prime\prime}l^{\prime};ABcd}(k^{\prime\prime},k^{\prime};\omega),\label{eq:4VC-red-irred}
\end{align}
and $\mathcal{J}^{(1)}_{ll^{\prime};\{a\}}(k,k^{\prime};\omega)$ is obtained 
by the method explained in Sect. 2.3, 
we can exactly calculate $\sigma_{\nu\nu}$ from Eqs. (\ref{eq:sigmaxx}), 
and (\ref{eq:sum-analytic}){--}(\ref{eq:4VC-red-irred}) in principle. 

Then, 
to rewrite Eq. (\ref{eq:sum-analytic}) in a more compact form, 
we use the three-point vector vertex function 
instead of the four-point vertex function. 
The three-point vertex function in Matsubara-frequency representation
is defined as
\begin{align}
&\textstyle\sum\limits_{A,B}
G_{aA}(\boldk+\boldq,i\omega_{m+n})
\Lambda_{\nu;AB}(\boldk,i\omega_{m};\boldq,i\Omega_{n})
G_{Bb}(\boldk,i\omega_{m})\notag\\
=&
\int^{T^{-1}}_{0}\hspace{-6pt}d\tau e^{i\omega_{m+n}\tau}
\int^{T^{-1}}_{0}\hspace{-6pt}d\tau^{\prime} e^{-i\Omega_{n}\tau^{\prime}}
\langle \textrm{T}_{\tau}  
\hat{c}_{\boldk+\boldq a}(\tau)
\hat{J}_{-\boldq \nu}(\tau^{\prime})
\hat{c}_{\boldk b}^{\dagger}
\rangle.
\end{align}
Since the analytic continuation of the three-point vector vertex function 
can be carried out\cite{NA-full} 
in a similar way to that used for $K_{abcd}(\boldk,\boldk^{\prime};i\Omega_{n})$, 
we obtain the three-point vector vertex function in real-frequency representation, 
\begin{align}
\Lambda_{\nu;l;ab}(k;q)
=&\
(v_{\boldk \nu})_{ab}
+\frac{1}{N}
\textstyle\sum\limits_{\boldk^{\prime}}
\textstyle\sum\limits_{\{A\}}
\textstyle\sum\limits_{l^{\prime}=1}^{3}
\int^{\infty}_{-\infty}\frac{d\epsilon^{\prime}}{4\pi i}
\mathcal{J}_{ll^{\prime};abCD}(k,k^{\prime};q)\notag\\
&\ \ \ \ \ \ \ \ \ \ \ \ \times 
g_{l^{\prime};CABD}(k^{\prime};q)
(v_{\boldk^{\prime} \nu})_{AB}.\label{eq:3-point-VC}
\end{align}
Here 
the additional subscript of the three-point vector vertex function, $l$, 
denotes the conditions about its $\epsilon$ and $\omega$: 
$l=1$ denotes $\textrm{Im}\epsilon > 0$ and 
$\textrm{Im}\epsilon+\textrm{Im}\omega > 0$; 
$l=2$ denotes $\textrm{Im}\epsilon < 0$ and 
$\textrm{Im}\epsilon+\textrm{Im}\omega > 0$; 
$l=3$ denotes $\textrm{Im}\epsilon < 0$ and 
$\textrm{Im}\epsilon+\textrm{Im}\omega < 0$. 
Combining Eqs. (\ref{eq:sum-analytic}) and (\ref{eq:Kpart-analytic}) 
with Eq. (\ref{eq:3-point-VC}), 
we can rewrite $\tilde{K}_{\nu \nu}^{(\textrm{R})}(\boldzero,\omega)$ as 
\begin{align}
\tilde{K}_{\nu \nu}^{(\textrm{R})}(\boldzero,\omega)
=&
-\frac{1}{N}
\textstyle\sum\limits_{\boldk}
\textstyle\sum\limits_{\{a\}}
(v_{\boldk \nu})_{ba}
\int^{\infty}_{-\infty}\frac{d\epsilon}{4\pi i}\notag\\
&\times \Bigl[
\tanh \frac{\epsilon}{2T}
g_{1;acdb}(k;\omega)\Lambda_{\nu;1;cd}(k;\omega)\notag\\
&\ \ \ +\Bigl(\tanh \frac{\epsilon+\omega}{2T}
-\tanh \frac{\epsilon}{2T}\Bigr)
g_{2;acdb}(k;\omega)\Lambda_{\nu;2;cd}(k;\omega)\notag\\
&\ \ \ -\tanh \frac{\epsilon+\omega}{2T}
g_{3;acdb}(k;\omega)\Lambda_{\nu;3;cd}(k;\omega)\Bigr].\label{eq:sum-analytic2}
\end{align} 

Since the exact expression is difficult to solve, 
we use the most-divergent-term approximation\cite{Eliashberg}, 
based on the properties\cite{Nozieres,AGD} of 
the product of the single-particle Green's functions in the limit 
$q\equiv (\boldq,\omega)\rightarrow 0$ 
in the presence of the QP peak.  
This is correct in the FL 
and remains appropriate in a correlated metallic system 
where perturbation expansion is satisfactory. 
As explained in Sect. 1, 
applicability of the FL theory differs from 
that of a microscopic perturbation theory. 
The microscopic perturbation theory is satisfactory 
to describe electronic properties of a correlated electron system 
if perturbation expansion has a good convergence 
or becomes an asymptotic expansion. 
Let us recall, first, that 
a well approximate partial sum can be constructed 
even if its convergence is not good\cite{Arfken-Weber}; 
second, that 
perturbation expansion becomes an asymptotic expansion near a phase transition. 

Before proceeding with the formal derivation of $\sigma_{\nu\nu}$, 
we remark on the property\cite{Nozieres,AGD} of 
a pair of the single-particle Green's functions in the limit $q\rightarrow 0$. 
When the QP peak exists and the QP damping is much smaller than $T$ (i.e., cold-spot-type), 
only $G_{ac}^{(\textrm{R})}(\boldk+\frac{\boldq}{2},\epsilon+\frac{\omega}{2})
G_{db}^{(\textrm{A})}(\boldk-\frac{\boldq}{2},\epsilon-\frac{\omega}{2})$ 
gives the most singular term being inversely proportional to the QP damping 
due to the merging of the poles of these Green's functions in $q\rightarrow 0$; 
the others, the retarded-retarded and the advanced-advanced pair, 
become the higher order terms. 
In this discussion, 
we have considered only the contribution from 
the coherent part of the single-particle Green's function 
since the incoherent part does not lead to such singular term in $q\rightarrow 0$. 
Also, 
I have used $\lim_{\delta\rightarrow 0+}[
\frac{1}{x-a+i\delta}-\frac{1}{x-a-i\delta}]=-2\pi i\delta(x-a)$ 
since  
the QP damping at momenta near the Fermi level 
is assumed to be negligible at low temperatures. 
The singular contribution from the hot spot is less important than 
that from the cold spot 
since the QP damping at the hot spot remains non-negligible even on the Fermi level. 
Since the existence of the QP peak and cold-spot-type QP damping 
is appropriate at least for several momenta in a metallic phase 
near a magnetic QCP\cite{ARPES-x02} or a Mott transition\cite{ARPES-Ir327}, 
the leading terms in $g_{l;acdb}(k; \omega)$ with respect to the QP damping or $\omega$ 
at low temperatures and frequencies are given by
\begin{align}
g_{1;acdb}(k;\omega)
\sim &
\textstyle\sum\limits_{\alpha,\beta}
u_{a\alpha c; d \beta b}(\boldk)
\dfrac{z_{\alpha}(\boldk)z_{\beta}(\boldk)}
{[\epsilon-\xi_{\alpha}^{\ast}(\boldk)+i0+]
[\epsilon-\xi_{\beta}^{\ast}(\boldk)+i0+]},\label{eq:g1-FL}\\
g_{2;acdb}(k;\omega)
\sim &\
2\pi i
\textstyle\sum\limits_{\alpha,\beta}
u_{a\alpha c; d \beta b}(\boldk)
\dfrac{z_{\alpha}(\boldk)z_{\beta}(\boldk)\delta(\epsilon-\xi_{\alpha}^{\ast}(\boldk))}
{\omega-\xi_{\alpha}^{\ast}(\boldk)+\xi_{\beta}^{\ast}(\boldk)
+i[\gamma_{\alpha}^{\ast}(\boldk)+\gamma_{\beta}^{\ast}(\boldk)]},\label{eq:g2-FL}
\end{align}
and
\begin{align}
\hspace{-50pt}
g_{3;acdb}(k;\omega)
\sim &
\textstyle\sum\limits_{\alpha,\beta}
u_{a\alpha c; d \beta b}(\boldk)
\dfrac{z_{\alpha}(\boldk)z_{\beta}(\boldk)}
{[\epsilon-\xi_{\alpha}^{\ast}(\boldk)-i0+]
[\epsilon-\xi_{\beta}^{\ast}(\boldk)-i0+]},\label{eq:g3-FL}
\end{align}
where $u_{a\alpha c; d \beta b}(\boldk)\equiv (U_{\boldk})_{a\alpha}(U_{\boldk}^{\dagger})_{\alpha c}
(U_{\boldk})_{d\beta}(U_{\boldk}^{\dagger})_{\beta b}$ with 
$(U_{\boldk})_{a\alpha}$ being the unitary matrix to obtain the QP bands 
[not equal to $(U_{\boldk}^{0})_{a\alpha}$]. 
Thus, 
the most divergent terms in the coherent limit arise from 
$g_{2;acdb}(k;\omega)$ at cold spots. 

Using the above property, 
we obtain an approximate expression of $\sigma_{\nu\nu}$ 
where we consider 
only the most divergent terms with respect to the QP damping in the coherent limit. 
To use the most-divergent-term approximation\cite{Eliashberg}, 
we introduce two quantities, 
$\mathcal{J}^{(0)}_{ll^{\prime};\{a\}}(k,k^{\prime};\omega)$ 
and $\Lambda_{\nu;l;ab}^{(0)}(k;\omega)$, 
which are irreducible with respect to only 
a retarded-advanced pair 
of the single-particle Green's functions, 
\begin{align}
\mathcal{J}^{(0)}_{ll^{\prime};\{a\}}(k,k^{\prime};\omega)
=&\ \mathcal{J}^{(1)}_{ll^{\prime};\{a\}}(k,k^{\prime};\omega) 
+\textstyle\sum\limits_{l^{\prime\prime}\neq 2}
\dfrac{1}{N}
\textstyle\sum\limits_{\boldk^{\prime\prime}}
\textstyle\sum\limits_{\{A\}}
\int^{\infty}_{-\infty}\frac{d\epsilon^{\prime\prime}}{4\pi i}
\mathcal{J}^{(0)}_{ll^{\prime\prime};abCD}(k,k^{\prime\prime};\omega)\notag\\
&\ \ \ \ \ \ \ \ \ \ \ \ \ \ \ \ \ \ \ \ \ \ \ \times
g_{l^{\prime\prime};CABD}(k^{\prime\prime};\omega)
\mathcal{J}^{(1)}_{l^{\prime\prime}l^{\prime};ABcd}(k^{\prime\prime},k^{\prime};\omega),\label{eq:4VC-0}
\end{align}
and
\begin{align}
\Lambda_{\nu;l;ab}^{(0)}(k;\omega)
=&\ 
(v_{\boldk \nu})_{ab}
+
\textstyle\sum\limits_{\{A\}}
\textstyle\sum\limits_{l^{\prime}\neq 2}
\dfrac{1}{N}\textstyle\sum\limits_{\boldk^{\prime}}
\int^{\infty}_{-\infty}\frac{d\epsilon^{\prime}}{4\pi i}
\mathcal{J}^{(0)}_{ll^{\prime};abCD}(k,k^{\prime};\omega)\notag\\
&\ \ \ \ \ \ \ \ \ \ \ \ \ \times
g_{l^{\prime};CABD}(k^{\prime};\omega)
(v_{\boldk^{\prime} \nu})_{AB}.\label{eq:3VC-0}
\end{align}
Using these two quantities 
with Eqs. (\ref{eq:4VC-12}), (\ref{eq:4VC-21}), 
(\ref{eq:4VC-23}) and (\ref{eq:4VC-32}) 
and the exchange symmetry of the four-point vertex function 
among its momentum and frequency variables, 
we can rewrite Eq. (\ref{eq:sum-analytic2}) as 
\begin{align}
\tilde{K}_{\nu \nu}^{(\textrm{R})}(\boldzero,\omega)
=&
-\frac{1}{N}
\textstyle\sum\limits_{\boldk}
\textstyle\sum\limits_{\{a\}}
(v_{\boldk \nu})_{ba}
\int^{\infty}_{-\infty}\dfrac{d\epsilon}{4\pi i}
\Bigl[\tanh \dfrac{\epsilon}{2T}
g_{1;acdb}(k;\omega)\Lambda_{\nu;1;cd}^{(0)}(k;\omega)\notag\\
&\ \ \ \ \ \ \ \ \ \ \ \ \ \ \ \ \ \ \ \ \ \ \ \ \ \ \ \ \ \ \ \ \ \ \ \ 
-\tanh \dfrac{\epsilon+\omega}{2T}
g_{3;acdb}(k;\omega)\Lambda_{\nu;3;cd}^{(0)}(k;\omega)\Bigr]\notag\\
& -\frac{1}{N}
\textstyle\sum\limits_{\boldk}
\textstyle\sum\limits_{\{a\}}
\Lambda_{\nu;2;ba}^{(0)}(k;\omega)
\int^{\infty}_{-\infty}\dfrac{d\epsilon}{4\pi i}
\Bigl(\tanh \dfrac{\epsilon+\omega}{2T}
-\tanh \dfrac{\epsilon}{2T}\Bigr)\notag\\
&\ \ \times g_{2;acdb}(k;\omega)\Lambda_{\nu;2;cd}(k;\omega).\label{eq:sum-analytic3}
\end{align}
At this stage, 
this expression remains exact. 
Then, 
since only the second term in Eq. (\ref{eq:sum-analytic3}) 
contains a retarded-advanced pair 
and 
the leading term with respect to $\omega$ comes from 
$(\tanh \frac{\epsilon+\omega}{2T}-\tanh \frac{\epsilon}{2T})
\approx 2\omega (-\frac{\partial f(\epsilon)}{\partial \epsilon})$, 
we can obtain an approximate expression of $\sigma_{\nu\nu}$ 
in the most-divergent-term approximation, 
\begin{align}
\sigma_{\nu\nu}=
\dfrac{2}{N}
\textstyle\sum\limits_{\boldk}
\textstyle\sum\limits_{\{a \}=1}^{3}
\int^{\infty}_{-\infty}\dfrac{d\epsilon}{2\pi}
\Lambda_{\nu;2;ba}^{(0)}(k;0)
g_{2;acdb}(k;0)
\Lambda_{\nu;2;cd}(k;0)
\Bigl(-\dfrac{\partial f(\epsilon)}{\partial \epsilon}\Bigr).\label{eq:sigmaxx-approx}
\end{align}
From this and Eq. (\ref{eq:g2-FL}), 
we can show that 
$\sigma_{\nu\nu}$ is inversely proportional to 
the unrenormalized QP damping\cite{Yamada-Yosida}. 

$\Lambda_{\nu;2;ba}^{(0)}(k;0)$ and $\Lambda_{\nu;2;cd}(k;0)$ in Eq. (\ref{eq:sigmaxx-approx}) 
are determined as follows. 
By combining Eq. (\ref{eq:3VC-0}) with the Ward identity\cite{Nozieres}, 
$\Lambda_{\nu;2;ba}^{(0)}(k;0)$ is given by 
\begin{align}
\Lambda_{\nu;2;ba}^{(0)}(k;0)=
(v_{\boldk \nu})_{ba}
+
\dfrac{\partial \textrm{Re}\Sigma_{ba}^{(\textrm{A})}(k)}{\partial k_{\nu}}.\label{eq:Lamb0}
\end{align}
In the present model, 
$\textrm{Re}\Sigma_{ba}^{(\textrm{A})}(k)=\textrm{Re}\Sigma_{ba}^{(\textrm{R})}(k)$ 
holds due to the even-parity and the time-reversal symmetry. 
Since 
$\mathcal{J}^{(0)}_{22;\{a\}}(k,k^{\prime};0)=\mathcal{J}^{(1)}_{22;\{a\}}(k,k^{\prime};0)$ 
is satisfied due to Eqs. (\ref{eq:4VC-12}), (\ref{eq:4VC-32}), and (\ref{eq:4VC-0}), 
$\Lambda_{\nu;2;cd}(k;0)$ is given by 
\begin{align}
\Lambda_{\nu;2;cd}(k;0)
=&\
\Lambda_{\nu;2;cd}^{(0)}(k;0)
+\frac{1}{N}
\textstyle\sum\limits_{\boldk^{\prime}}
\textstyle\sum\limits_{\{A\}}
\int^{\infty}_{-\infty}\frac{d\epsilon^{\prime}}{4\pi i}
\mathcal{J}_{22;cdCD}^{(1)}(k,k^{\prime};0)
\notag\\
&\ \ \ \ \ \ \ \ \ \ \ \ \ \ \ \ \ \ \times 
g_{2;CABD}(k^{\prime};0)
\Lambda_{\nu;2;AB}(k^{\prime};0).\label{eq:Lamb}
\end{align}
Thus, 
the present theory takes account of the CVCs 
due to the self-energy and irreducible four-point vertex function, 
neglected in the relaxation-time approximation\cite{Ziman}. 

\subsection{FLEX approximation with the CVCs 
arising from the self-energy and MT term for a multiorbital Hubbard model in a PM phase}
To calculate the resistivity, 
I use the FLEX approximation 
with the CVCs arising from the self-energy and MT term in a PM phase.   
I first explain the FLEX approximation for a multiorbital Hubbard model in a PM phase. 
Next, I derive the irreducible four-point vertex function in this approximation; 
as described in Sect. 1, 
I consider only the MT term and neglect the AL term. 
Then, I show the Bethe-Salpeter equation for the current 
by using the MT term as the kernel of the CVC. 
I also remark on the numerical treatment of the principal integral 
appearing in that CVC. 
Finally, I discuss applicability of this method. 
 
I determine several single-particle or two-particle quantities 
using the FLEX approximation\cite{FLEX1,FLEX2,multi-FLEX} in a PM phase 
where only the electron-hole scattering processes of the bubble and the ladder diagrams 
are considered as the Luttinger-Ward functional\cite{Luttinger-Ward,Baym-Kadanoff}, 
$\Phi_{\textrm{LW}}[G]$. 
Since that is a conserving approximation\cite{Luttinger-Ward,Baym-Kadanoff} 
based on the thermodynamic potential expressed in terms of the single-particle Green's function, 
we can determine single-particle or two-particle quantities 
by using $\Phi_{\textrm{LW}}[G]$ and its functional derivative. 
Since constructing $\Phi_{\textrm{LW}}[G]$ is equivalent to 
constructing the effective interaction, 
where the same kinds of the diagrams are considered, 
and the latter is easier, 
we formulate the FLEX approximation by the latter procedure as follows. 
First, considering the electron-hole scattering processes 
of the bubble and the ladder diagrams for $\hat{H}_{\textrm{int}}$, 
we obtain the effective interaction in the FLEX approximation, 
\begin{align}
\hspace{-4pt}
V_{abcd}^{s_{1}s_{2}s_{3}s_{4}}(\boldq,i\Omega_{n})
=&\ 
\frac{1}{2}
\Bigl[
U_{abcd}^{\textrm{C}}
-\textstyle\sum\limits_{\{A\}}
U_{abAB}^{\textrm{C}}\chi_{ABCD}^{\textrm{C}}(\boldq,i\Omega_{n})U_{CDcd}^{\textrm{C}}
\Bigr]\sigma_{s_{1} s_{2}}^{0}\sigma_{s_{4} s_{3}}^{0}\notag\\
\hspace{-4pt}
-&\frac{1}{2}
\Bigl[
U_{abcd}^{\textrm{S}}
+\textstyle\sum\limits_{\{A\}}
U_{abAB}^{\textrm{S}}\chi_{ABCD}^{\textrm{S}}(\boldq,i\Omega_{n})U_{CDcd}^{\textrm{S}}
\Bigr]\boldsigma_{s_{1} s_{2}}\cdot \boldsigma_{s_{4} s_{3}},
\end{align}
with 
\begin{align}
\chi_{abcd}^{\textrm{S}}(\boldq,i\Omega_{n})
=&\ 
\chi_{abcd}(\boldq,i\Omega_{n})
+\textstyle\sum\limits_{\{A\}}
\chi_{abAB}(\boldq,i\Omega_{n})
U_{ABCD}^{\textrm{S}}
\chi_{CDcd}^{\textrm{S}}(\boldq,i\Omega_{n}),\label{eq:FLEX-1}\\
\chi_{abcd}^{\textrm{C}}(\boldq,i\Omega_{n})
=&\ 
\chi_{abcd}(\boldq,i\Omega_{n})
-\textstyle\sum\limits_{\{A\}}
\chi_{abAB}(\boldq,i\Omega_{n})
U_{ABCD}^{\textrm{C}}
\chi_{CDcd}^{\textrm{C}}(\boldq,i\Omega_{n}),\label{eq:FLEX-2}
\end{align}
and
\begin{align}
\hspace{-73pt}
\chi_{abcd}(\boldq,i\Omega_{n})
=&-
\frac{T}{N}
\textstyle\sum\limits_{\boldk}
\textstyle\sum\limits_{m}
G_{ac}(\boldk+\boldq,i\omega_{m+n})
G_{db}(\boldk,i\omega_{m}).\label{eq:FLEX-3}
\end{align}
Here we introduce the bare four-point vertex functions 
in spin and charge sector,
\begin{align}
& U_{abcd}^{\textrm{S}}
=\ U^{\uparrow\downarrow}_{abcd}-U^{\uparrow\uparrow}_{abcd} 
=
\begin{cases} 
\ U \ \ \ \ \ \ \ \ \ \  \textrm{for} \ a=b=c=d\\
\ J_{\textrm{H}} \ \ \ \ \ \ \ \ \ \textrm{for} \ a=b\neq c=d\\
\ U^{\prime} \ \ \ \ \ \ \  \  \ \textrm{for} \ a=c\neq b=d\\
\ J^{\prime}\ \ \ \ \ \ \ \ \ \ \textrm{for} \ a=d\neq b=c\\
\end{cases},\label{eq:bareS}
\end{align}
and
\begin{align}
& U_{abcd}^{\textrm{C}}
=\ U^{\uparrow\downarrow}_{abcd}+U^{\uparrow\uparrow}_{abcd} 
=
\begin{cases} 
\ U \ \ \ \ \ \ \ \ \ \ \ \  \  \ \textrm{for} \ a=b=c=d\\
\ 2U^{\prime}-J_{\textrm{H}} \  \ \ \ \ \textrm{for} \ a=b\neq c=d\\
\ -U^{\prime}+2J_{\textrm{H}} \ \ \ \textrm{for} \ a=c\neq b=d\\
\ J^{\prime}\ \ \ \ \ \ \ \ \ \ \ \ \ \ \textrm{for} \ a=d\neq b=c\\
\end{cases},\label{eq:bareC}
\end{align}
where $U_{abcd}^{s s^{\prime}}$ is 
$U_{abcd}^{s s^{\prime}}\equiv U^{s s s^{\prime} s^{\prime}}_{abcd}$ 
and the spin-flipping term, $U^{ss^{\prime}ss^{\prime}}_{abcd}$ for $s\neq s^{\prime}$, satisfies 
$U^{\uparrow \downarrow \uparrow \downarrow}_{abcd}=-U_{acbd}^{\uparrow \downarrow}=-U_{abcd}^{\textrm{S}}$. 
Also, 
we neglect the vertex corrections to the susceptibilities in spin and charge sector; 
its effects are discussed later. 
Then, 
the single-particle Green's function is determined by the Dyson equation, 
\begin{align}
G_{ab}(\boldk,i\omega_{m})
=&\ G_{ab}^{0}(\boldk,i\omega_{m})
+\textstyle\sum\limits_{A,B}
G_{aA}^{0}(\boldk,i\omega_{m})
\Sigma_{AB}(\boldk,i\omega_{m})
G_{Bb}(\boldk,i\omega_{m}),\label{eq:FLEX-4}
\end{align}
with the self-energy given by
\begin{align}
\Sigma_{ac}(\boldk,i\omega_{m})
=&
\frac{T}{N}
\textstyle\sum\limits_{\boldq}
\textstyle\sum\limits_{n}
\textstyle\sum\limits_{b,d}
V_{abcd}(\boldq,i\Omega_{n})
G_{bd}(\boldk-\boldq,i\omega_{m-n}),\label{eq:FLEX-5}
\end{align}
where 
\begin{align}
V_{abcd}(\boldq,i\Omega_{n})
=&
-V_{abcd}^{\uparrow\uparrow\uparrow\uparrow}(\boldq,i\Omega_{n})
-V_{abcd}^{\uparrow\downarrow\uparrow\downarrow}(\boldq,i\Omega_{n})
-\textstyle\sum\limits_{\{A\}}
U_{aAbB}^{\uparrow\downarrow}\chi_{ABCD}(\boldq,i\Omega_{n})U_{CcDd}^{\uparrow\downarrow}\notag\\
=&\
\dfrac{3}{2}
\Bigl[
U_{abcd}^{\textrm{S}}
+\textstyle\sum\limits_{\{A\}}
U_{abAB}^{\textrm{S}}
\chi_{ABCD}^{\textrm{S}}(\boldq,i\Omega_{n})
U_{CDcd}^{\textrm{S}}\Bigr]\notag\\
&
+\dfrac{1}{2}
\Bigl[
-U_{abcd}^{\textrm{C}}
+\textstyle\sum\limits_{\{A\}}
U_{abAB}^{\textrm{C}}
\chi_{ABCD}^{\textrm{C}}(\boldq,i\Omega_{n})
U_{CDcd}^{\textrm{C}}
\Bigr]\notag\\
&
-\textstyle\sum\limits_{\{A\}}
U_{aAbB}^{\uparrow\downarrow}
\chi_{ABCD}(\boldq,i\Omega_{n})
U_{CcDd}^{\uparrow\downarrow}.\label{eq:FLEX-6}
\end{align}
The last term in Eq. (\ref{eq:FLEX-6}) 
is introduced to exclude the double counting of the topologically equivalent term 
in the self-energy. 
Solving Eqs. (\ref{eq:FLEX-1}){--}(\ref{eq:FLEX-3}) and (\ref{eq:FLEX-4}){--}(\ref{eq:FLEX-6}) 
with Eqs. (\ref{eq:mu-int}), (\ref{eq:G0}), (\ref{eq:bareS}) and (\ref{eq:bareC})  
selfconsistently by iteration, 
we can determine the single-particle and the two-particle quantities in the FLEX approximation. 

It should be noted that 
the partial inclusion of mode-mode couplings\cite{Moriya-review} for fluctuations, 
the interactions between fluctuations at different momenta, 
through the self-energy 
improves some unrealistic results obtained in the random-phase approximation, 
although the susceptibilities are determined by 
the random-phase approximation-type (but renormalized) equations. 
For example, 
in the FLEX approximation\cite{Kon-CVC,NA,multi-FLEX,Yanase-review},
the value of $U$ for a magnetic transition becomes about $2$ eV, 
the momentum dependences of the mass enhancement and FS deformation 
are taken into account, 
and the Curie-Weiss-like temperature dependence of the spin susceptibility is obtained 
near a magnetic QCP. 
In particular, the final improvement 
is powerful to describe electronic properties near a magnetic QCP. 

We also determine the irreducible four-point vertex function 
in the FLEX approximation in keeping conservation laws\cite{FLEX2}. 
In a conserving approximation, 
the irreducible four-point vertex function is given by\cite{Kon-CVC,FLEX2,Baym-Kadanoff} 
\begin{align}
\Gamma_{abcd}^{(1)}(\boldk,i\omega_{m},\boldk^{\prime},i\omega_{m^{\prime}}; \boldq,i\Omega_{n})
=\ 
\dfrac{\delta \Sigma_{ab}(\boldk,i\omega_{m})}
{\delta G_{cd}(\boldk^{\prime},i\omega_{m^{\prime}})}.\label{eq:Gamma1-consv}
\end{align}
For the actual calculations, 
we first calculate the right-hand side at $\boldq=\boldzero$ and $\Omega_{n}=0$ 
and then label momentum and frequency transfers correctly 
as the electron-hole scattering process among 
an electron of orbital $b$ with $(\boldk,i\omega_{m})$, 
a hole of orbital $d$ with $(\boldk^{\prime},i\omega_{m^{\prime}})$, 
an electron of orbital $a$ with $(\boldk+\boldq,i\omega_{m+n})$, 
and a hole of orbital $c$ with $(\boldk^{\prime}+\boldq,i\omega_{m^{\prime}+n})$. 
After several straightforward calculations\cite{NA-full} 
by using Eqs. (\ref{eq:FLEX-5}){--}(\ref{eq:Gamma1-consv}), 
we obtain the irreducible four-point vertex function in the FLEX approximation, 
which is the sum of the MT and the AL term\cite{NA-full}. 
In this paper, 
I consider only the MT term,
\begin{align}
\Gamma_{abcd}^{(1)}(\boldk,i\omega_{m},\boldk^{\prime},i\omega_{m^{\prime}}; 
\boldq,i\Omega_{n})
=& \ V_{acbd}(\boldk-\boldk^{\prime},i\omega_{m}-i\omega_{m^{\prime}}).\label{eq:MT-ImFreq}
\end{align}
This treatment will be sufficient 
for a qualitative discussion about many-body effects on $\rho_{ab}$ 
since the CVC arising from the AL term gives the higher order contribution compared with 
that arising from the MT term\cite{NA}. 
I have checked the validity of this statement 
by calculating the main terms of the AL CVC\cite{NA-full}.
Since it is necessary to calculate 
$\mathcal{J}^{(1)}_{22;cdCD}(k,k^{\prime};0)$ in Eq. (\ref{eq:Lamb}), 
we need to carry out the analytic continuation of 
Eq. (\ref{eq:MT-ImFreq}) 
in region $22$-II, region $22$-III, and region $22$-IV.  
Carrying out the analytic continuation\cite{NA-full} of the MT term 
and using Eq. (\ref{eq:4VC-22}), we obtain 
\begin{align}
\mathcal{J}_{22;cdCD}^{(1)}(k,k^{\prime};0)
=\  
2i \Bigl(\coth \frac{\epsilon-\epsilon^{\prime}}{2T}
+\tanh \frac{\epsilon^{\prime}}{2T}
 \Bigr) \textrm{Im}V_{cCdD}^{(\textrm{R})}(k-k^{\prime}).\label{eq:Gamma1-FLEX-ReFreq}
\end{align}

Substituting Eq. (\ref{eq:Gamma1-FLEX-ReFreq}) into Eq. (\ref{eq:Lamb}), 
we obtain the Bethe-Salpeter equation for the current 
in the FLEX approximation with the MT CVC in a PM phase,
\begin{align}
\Lambda_{\nu;2;cd}(k;0)
=&\ \Lambda_{\nu;2;cd}^{(0)}(k)
+\frac{1}{N}
\textstyle\sum\limits_{\boldk^{\prime}}
\textstyle\sum\limits_{\{A \}}
\int^{\infty}_{-\infty}\dfrac{d\epsilon^{\prime}}{2\pi}
\Bigl(\coth \dfrac{\epsilon-\epsilon^{\prime}}{2T}
+\tanh \dfrac{\epsilon^{\prime}}{2T}
 \Bigr)\notag\\
&\ \ \ \ \ \ \ \ \ \ \ \ \ \ \ \times 
\textrm{Im}V_{cCdD}^{(\textrm{R})}(k-k^{\prime})
g_{2;CABD}(k^{\prime};0)
\Lambda_{\nu;2;AB}(k^{\prime};0).\label{eq:BS-MTCVC}
\end{align} 
We see the roles of the MT CVC are similar to 
those of the backflow correction\cite{Nozieres}. 

Before discussing applicability, 
I explain how to treat the principal integral 
in Eq. (\ref{eq:BS-MTCVC}) for the numerical calculations. 
Since both the numerator and denominator of the term containing 
$\coth \frac{\epsilon-\epsilon^{\prime}}{2T}$ in the MT CVC 
become zero simultaneously at $\epsilon^{\prime}=\epsilon$ 
due to $\textrm{Im}V_{abcd}^{(\textrm{R})}(\boldq,0)=0$, 
the principal integral can be calculated as follows: 
\begin{align}
&
\int^{\infty}_{-\infty}\frac{d\epsilon^{\prime}}{2\pi}
\coth \frac{\epsilon-\epsilon^{\prime}}{2T}
\textrm{Im}V_{cCdD}^{(\textrm{R})}(k-k^{\prime})
g_{2;CABD}(k^{\prime};0)
\Lambda_{\nu;2;AB}(k;0)\notag\\
=&\ 
\int_{\epsilon^{\prime}\neq \epsilon}\frac{d\epsilon^{\prime}}{2\pi}
\coth \frac{\epsilon-\epsilon^{\prime}}{2T}
\textrm{Im}V_{cCdD}^{(\textrm{R})}(k-k^{\prime})
g_{2;CABD}(k^{\prime};0)
\Lambda_{\nu;2;AB}(k;0)\notag\\
-&\dfrac{\Delta \epsilon^{\prime}}{2\pi}
T
\dfrac{\partial}
{\partial \epsilon^{\prime}}
\bigl[
(e^{\frac{\epsilon^{\prime}-\epsilon}{T}}+1)
\textrm{Im}V_{cCdD}^{(\textrm{R})}(k-k^{\prime})
g_{2;CABD}(k^{\prime};0)
\Lambda_{\nu;2;AB}(k^{\prime};0)\bigr]
\Bigl|_{\epsilon^{\prime}=\epsilon},
\end{align}
where the first term contains the contributions other than $\epsilon^{\prime}=\epsilon$.

Finally, I discuss applicability of the FLEX approximation with the CVCs 
arising from the self-energy and MT term. 

First, 
the FLEX approximation is suitable to describe the electronic structure 
at low temperatures for moderately strong electron correlation. 
In a single-orbital Hubbard model on a square lattice\cite{FLEX1,FLEX3}, 
the imaginary-time dependence of the single-particle Green's function 
at several momenta in the FLEX approximation 
shows satisfactory (but not perfect) agreement with 
that in the quantum-Monte-Carlo calculation 
at $U$ being a half of the bandwidth; 
the agreement becomes better near the AF QCP than away from the AF QCP. 
Since 
the similar agreement will hold even in a multiorbital Hubbard model on the same lattice 
and 
the FLEX approximation can treat 
the coherent part of the electronic spectrum satisfactorily\cite{Ikeda-4th}, 
the electronic structure in metallic phases of the present model at low temperatures 
will be well described by the FLEX approximation at least qualitatively. 
Actually, 
the FLEX approximation succeeded 
in reproducing the larger effective mass\cite{dHvA-x2} of the $d_{xy}$ orbital 
than that of the $d_{xz/yz}$ orbital\cite{NA} 
and its agreement with the experiment\cite{dHvA-x2} is better than 
the case of the DMFT\cite{Haule-DMFT}, as described in Sect. 1. 

In contrast, 
the FLEX approximation becomes unsuitable 
at high temperatures or near a Mott transition 
for strong electron correlation. 
This is because 
local correlation plays important roles in such case\cite{Georges-review,Kotliar-review} 
and 
the effects of local correlation on the electronic spectrum 
are smeared out in the FLEX approximation\cite{Ikeda-4th}. 

Then, the magnetic properties at low temperatures for moderately strong electron correlation 
will be appropriately described by the FLEX approximation 
if the dominant correlation of the system 
is spin fluctuation whose largest contribution comes from a non-degenerate orbital. 
Due to neglecting the vertex corrections to the susceptibilities, 
the enhancement of spin fluctuation arising from electron correlation 
is overestimated in the FLEX approximation 
compared with the enhancement of charge or orbital fluctuation. 
Actually, 
in a two-degenerate-orbital Hubbard model on a square lattice 
at small $(J_{\textrm{H}}/U)$ near an AF QCP, 
the AL vertex correction to the susceptibilities causes 
the enhancement of orbital fluctuation\cite{AL-2orb}. 
However, 
I believe that 
in the present model 
the FLEX approximation is sufficient to describe the magnetic properties at least qualitatively 
since the $d_{xy}$ orbital gives the largest contribution to spin fluctuation\cite{NA}; 
in this case, 
even if the MT and the AL vertex correction to the susceptibilities are considered 
beyond the FLEX approximation, 
the magnetic properties will not qualitatively change 
and 
these corrections will modify the values of the susceptibilities 
since orbital fluctuation enhanced due to the AL term 
does not dominate over spin fluctuation. 
Actually, 
the strongest enhancement of spin fluctuation at 
$\boldQ_{\textrm{IC-AF}}\equiv (\frac{21\pi}{32},\frac{21\pi}{32})$ 
in the FLEX approximation\cite{NA} away from and near the AF QCP 
agrees with the experiments in Sr$_{2}$RuO$_{4}$\cite{Neutron-x2} 
and Sr$_{2}$Ru$_{0.075}$Ti$_{0.025}$O$_{4}$\cite{Neutron-Ti214}, 
respectively. 

Moreover, 
if the vertex corrections to the current arising from 
the self-energy and irreducible four-point vertex function due to electron correlation 
are added to the FLEX approximation, 
this method is suitable to describe the transport properties of a metallic phase 
due to low-frequency external field satisfying $\omega \tau_{\textrm{trans}} \ll 1$ 
at low temperatures. 
In contrast to case of the vertex corrections to the susceptibilities, 
the vertex corrections to the current 
are essential for discussing the transport properties\cite{Nozieres,Yamada-Yosida} 
since the CVCs are vital to satisfy conservation laws 
and conservation laws play significant roles in transport phenomena. 
For example, 
the importance of the treatment holding conservation laws 
is known for a system without the lattice (e.g., the electron gas): 
only if the CVCs due to electron correlation are correctly taken into account,  
we can obtain the correct results such as 
the absence both of the resistivity\cite{Yamada-Yosida} and 
of the renormalization of the Drude weight\cite{Maebashi} and 
electron cyclotron frequency\cite{Kohn-thm}. 
Also, 
the CVCs due to electron correlation are important in a system with the lattice 
since these CVCs are necessary to obtain the correct effects of electron correlation 
on the transport coefficients in the presence of 
the Umklapp scattering\cite{Yamada-Yosida,Maebashi,Kanki-Yamada}. 
Another example showing the importance of the CVCs 
is the emergence of the Curie-Weiss-like temperature dependence of 
the Hall coefficient near the AF QCP due to the MT CVC in the FLEX approximation 
in the single-orbital Hubbard model on a square lattice\cite{Kon-CVC}. 
In addition to the treatment holding conservation laws, 
the satisfactory treatment of the coherent part in the FLEX approximation 
is powerful to describe the transport properties of a metallic phase 
due to the low-frequency external field. 
This is because 
the dominant contributions to the response induced by that external field 
at low temperatures 
come from the contributions near the Fermi level 
as a result of the energy derivative of the Fermi distribution function 
in the response function. 
Furthermore, 
this powerfulness of the FLEX approximation will hold even near a Mott transition 
since it is shown 
in the DMFT\cite{DMFT-trans-QP} for a single-orbital Hubbard model 
that 
the transport properties in $\omega\tau_{\textrm{trans}} \ll 1$ 
are well described in the approximation where only the coherent part is considered. 
Thus, 
the FLEX approximation with the CVCs arising from the self-energy and MT term 
is satisfactory to describe the transport properties 
of metallic phases of the present model at low temperatures. 
Actually, 
this method\cite{NA} near and away from the AF QCP 
reproduced the temperature dependence of several transport properties 
of Sr$_{2}$RuO$_{4}$\cite{resistivity-x2,Hall-x2} 
and Sr$_{2}$Ru$_{0.075}$Ti$_{0.025}$O$_{4}$\cite{Ti214-nFL2}, 
as pointed out in Sect. 1. 

\section{Results}
In this section, 
I show the results of $\rho_{ab}=\sigma_{xx}^{-1}$ of some quasi-$2$D PM ruthenates  
in the FLEX approximation 
with the CVCs arising from the self-energy and MT term 
and compare these results obtained in more simplified cases than that method. 
In particular, 
we focus on the effects of the self-energy and MT term 
due to electron correlation and the role of each $t_{2g}$ orbital. 

\begin{figure}[tb]
\begin{center}
\includegraphics[width=125mm]{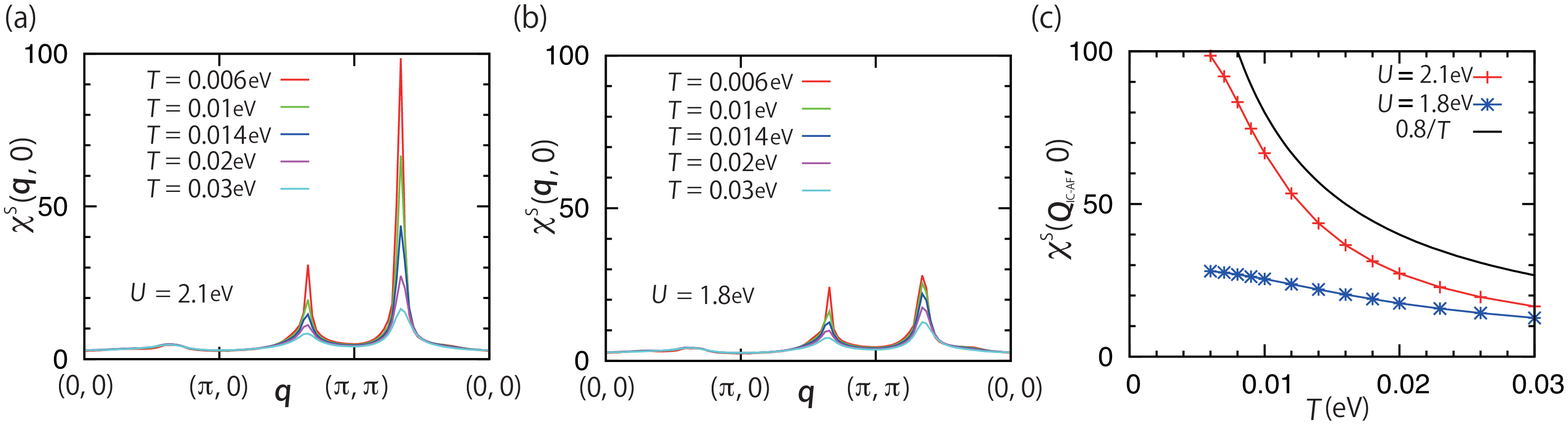}
\end{center}
\vspace{-18pt}
\caption{
Momentum dependences of the static spin susceptibility, 
$\chi^{\textrm{S}}(\boldq,0)=\sum_{a,b}\chi^{\textrm{S}}_{aabb}(\boldq,0)$, 
for several temperatures at $U=$ (a) $1.8$ and (b) $2.1$ eV, 
and (c) temperature dependences of $\chi^{\textrm{S}}(\boldQ_{\textrm{IC-AF}},0)$ 
at $U=1.8$ and $2.1$ eV. 
}
\label{fig:Fig3}
\end{figure}

I carried out the numerical calculations as follows. 
I set the $64\times 64$ meshes of the Brillouin zone 
and $2048$ Matsubara frequencies 
and used the fast Fourier transformation 
with the zero padding method\cite{NumericalReceipt}. 
I obtained the single-particle Green's function, self-energy, 
and MT term in the FLEX approximation 
by solving Eqs. (\ref{eq:FLEX-1}){--}(\ref{eq:FLEX-3}) 
and (\ref{eq:FLEX-4}){--}(\ref{eq:FLEX-6}) 
with Eqs. (\ref{eq:mu-int}), (\ref{eq:G0}), (\ref{eq:bareS}) and (\ref{eq:bareC}) 
by iteration, 
where I assumed that 
convergence was reached when the difference between the self-energies 
before and after certain iteration was less than $10^{-4}$. 
To obtain the quantities as a function of real frequency, 
I used the Pad\'{e} approximation\cite{Pade-approx} 
using the quantities at the lowest four Matsubara frequencies. 
The real-frequency integrations were approximated 
by the integrations with the interval $0.0025$ eV and 
the upper and lower values $0.2$ and $-0.2$ eV. 
The current was determined by solving Eq. (\ref{eq:BS-MTCVC}) by iteration, 
where the convergence condition was assumed to be that 
the difference between the currents before and after certain iteration 
was less than $10^{-4}$. 

In the following, 
I consider cases at $U=1.8$ and $2.1$ eV 
as cases away from and near the AF QCP, respectively. 
These correspondences are because of 
the Pauli-PM temperature dependence of the spin susceptibility at $U=1.8$ eV 
and the Curie-Weiss-like temperature dependence of the spin susceptibility 
at $\boldq=\boldQ_{\textrm{IC-AF}}$ at $U=2.1$ eV 
[see Figs. \ref{fig:Fig3}(a){--}\ref{fig:Fig3}(c)]. 
The latter is characteristic of a magnetic QCP 
and causes the strong temperature dependence of the Landau parameters 
with momentum transfer $\boldQ_{\textrm{IC-AF}}$ 
through the temperature dependence of the reducible four-point vertex function. 
The choice of $U$ is reasonable 
since the value estimated experimentally in Sr$_{2}$RuO$_{4}$ is about $2$ eV\cite{X-ray10Dq}. 

As explained below in detail, 
there are three main results: 
(i) $\rho_{ab}$ of some quasi-$2$D PM ruthenates 
without the rotation and the tilting of RuO$_{6}$ octahedra 
is determined almost by the conductions of the $d_{xz/yz}$ orbital 
due to the smaller unrenormalized QP dampings of the $d_{xz/yz}$ orbital 
than those of the $d_{xy}$ orbital; 
(ii) The crossover between the $T$-linear and the $T^{2}$ dependence of $\rho_{ab}$ 
occurs away from the AF QCP at about $T=0.008$ eV 
due to the temperature dependences of the unrenormalized QP dampings of the $d_{xz/yz}$ orbital; 
(iii) The $T$-linear $\rho_{ab}$ emerges near the AF QCP 
due to the hot-spot structure of the QP dampings of the $d_{xz/yz}$ orbital 
at $\boldk=\boldQ_{\textrm{IC-AF}}$ and its symmetrically equivalent ones.  

\subsection{Case away from the AF QCP}
\begin{figure}[tb]
\begin{center}
\includegraphics[width=72mm]{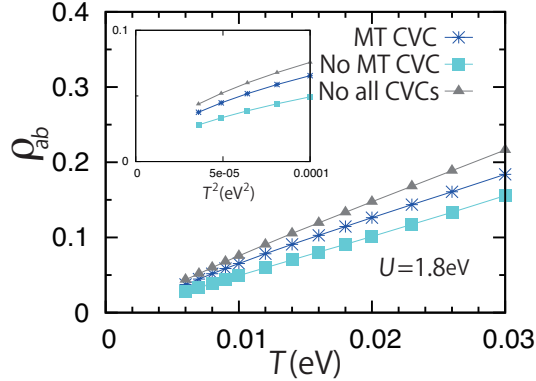}
\end{center}
\vspace{-18pt}
\caption{
Temperature dependence of $\rho_{ab}$ at $U=1.8$ eV.  
The definition of each case is described in the main text. 
The inset shows $\rho_{ab}$ against $T^{2}$ below $T=0.01$ eV.  
}
\label{fig:Fig4}
\end{figure}

I begin with the temperature dependence of $\rho_{ab}$ at $U=1.8$ eV in Fig. \ref{fig:Fig4}. 
Three cases in that figure are defined as follows: 
in MT CVC case, 
the CVCs arising from the self-energy and MT term are included; 
in No MT CVC case, the CVC arising from the self-energy is included; 
in No all CVCs case, 
which is equivalent to the relaxation-time approximation\cite{Ziman}, 
all the CVCs are neglected. 
There are three main remarks about Fig. \ref{fig:Fig4}. 
First, 
in all the three cases, 
decreasing temperature causes the crossover from the $T$-linear dependence 
to the $T^{2}$ dependence at about $T=0.008$ eV. 
The $T^{2}$ dependence at low temperatures 
is more clearly seen from the inset of Fig. \ref{fig:Fig4}. 
Second, 
the value of $\rho_{ab}$ in the No all CVCs case at each temperature
is largest in the three cases. 
Third, 
the value of $\rho_{ab}$ increases from that in the No MT CVC case 
when the MT CVC is included. 
The first remark indicates that
the CVCs little affect on 
the power of the temperature dependence of the resistivity. 
The second indicates that 
the value of the resistivity is overestimated in the relaxation-time approximation. 
The third indicates that 
the MT CVC enhances the resistivity as a result of the reduction of the current, 
which is similar to the effect of the backflow correction. 
Thus, 
the main effects of the CVCs on the resistivity just change its value. 

\begin{figure}[tb]
\begin{center}
\includegraphics[width=123mm]{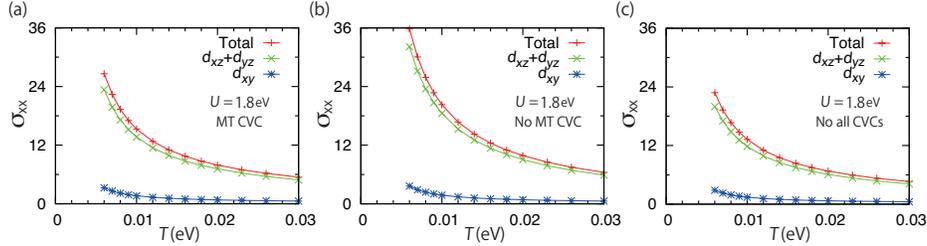}
\end{center}
\vspace{-18pt}
\caption{
Temperature dependences of the total and orbital-decomposed components of $\sigma_{xx}$ 
at $U=1.8$ eV in (a) the MT CVC case, (b) the No MT CVC case, 
and (c) the No all CVCs case.  
The definition of each orbital-decomposed $\sigma_{xx}$ is described in the main text.  
}
\label{fig:Fig5}
\end{figure}
\begin{figure}[tb]
\vspace{-10pt}
\includegraphics[width=123mm]{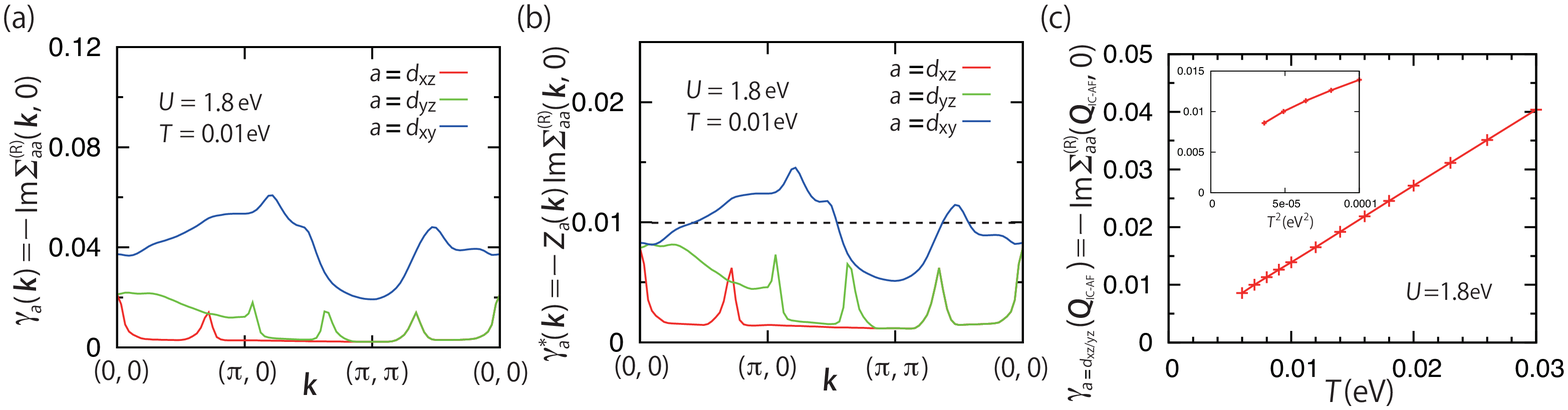}
\vspace{-12pt}
\caption{
Momentum dependence of (a) the unrenormalized QP damping and (b) the QP damping 
for each $t_{2g}$ orbital 
at $(U,T)=(1.8,0.01)$ (eV), 
and (c) temperature dependence of the unrenormalized QP damping of the $d_{xz/yz}$ orbital 
at $\boldk=\boldQ_{\textrm{IC-AF}}$ at $U=1.8$ eV. 
The dashed line in panel (b) denotes $T=0.01$ eV. 
The inset in panel (c) shows the data against $T^{2}$. 
}
\label{fig:Fig6}
\end{figure}

Then, 
to understand the role of each $t_{2g}$ orbital, 
I analyze the orbital-decomposed $\sigma_{xx}$ at $U=1.8$ eV. 
The orbital-decomposed $\sigma_{xx}$ for the $d_{xz/yz}$ and the $d_{xy}$ orbital 
are obtained by replacing 
$\sum_{\{a \}=1}^{3}$ in Eq. (\ref{eq:sigmaxx-approx}) 
by $\sum_{\{a \}=1}^{2}$ and $\sum_{\{a \}=3}$, respectively. 
Those components are sufficient in the present model 
since the intraorbital components are much larger than the interobital components 
due to the large intraorbital hopping integrals compared with the interorbital ones.  
We see from Figs. \ref{fig:Fig5}(a){--}\ref{fig:Fig5}(c) that 
the dominant contributions to the total of $\sigma_{xx}$ 
come from the component of the $d_{xz}$ orbital, 
and that 
the component of the $d_{xy}$ orbital 
is less than $10\%$ of the total. 
Due to the rotational symmetry of the system, 
in case of $\sigma_{yy}$, 
the component of the $d_{yz}$ orbital 
gives the dominant contributions. 
Thus, 
the inplane transport is governed mainly by 
the conductions of the $d_{xz/yz}$ orbital. 

The above orbital-dependent transport arises from 
the smaller unrenormalized QP dampings of the $d_{xz/yz}$ orbital 
than those of the $d_{xy}$ orbital 
since $\sigma_{\nu\nu}$ is inversely proportional to the unrenormalized QP damping, 
as explained in Sect. 2.2. 
Actually, 
we see from Fig. \ref{fig:Fig6}(a) that 
the unrenormalized QP dampings of the $d_{xz/yz}$ orbital are smaller. 
The similar orbital dependence holds at the other temperatures. 

Also, 
we see from Figs. \ref{fig:Fig6}(b) and \ref{fig:Fig6}(c), respectively, 
that 
the QP dampings of the $d_{xz/yz}$ orbital remain the cold spot, 
and that 
the temperature dependence of the unrenormalized QP damping of the $d_{xz/yz}$ orbital 
at $\boldk=\boldQ_{\textrm{IC-AF}}$ changes from $T^{2}$ to $T$-linear at about $T=0.008$ eV. 
Thus, 
the latter is the origin of 
the crossover of the power of the temperature dependence of $\rho_{ab}$ at about $T=0.008$ eV. 

\subsection{Case near the AF QCP}
\begin{figure}[tb]
\begin{center}
\includegraphics[width=72mm]{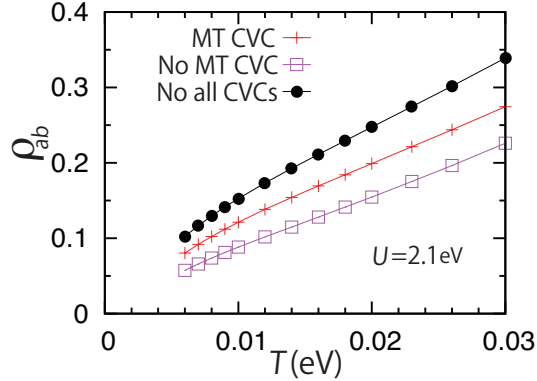}
\end{center}
\vspace{-16pt}
\caption{
Temperature dependence of $\rho_{ab}$ at $U=2.1$ eV.  
The definition of each case is described in the main text in Sect. 3.1.  
}
\label{fig:Fig7}
\end{figure}

I turn to the temperature dependence of $\rho_{ab}$ at $U=2.1$ eV 
in the three cases, considered in Sect. 3.1. 
From Fig. \ref{fig:Fig7}, 
we see that 
the $T$-linear $\rho_{ab}$ emerges in all the three cases. 
We also see the similar effects of the CVCs on the value of $\rho_{ab}$ 
to those at $U=1.8$ eV, 
the overestimation in the relaxation-time approximation (i.e., the No all CVCs case) 
and the increase from the value in the No MT CVC case due to the MT CVC.  
The first result indicates that 
the emergence of the $T$-linear $\rho_{ab}$ near the AF QCP 
arises from the temperature dependence of the unrenormalized QP damping. 
Furthermore, 
those results and the corresponding results at $U=1.8$ eV 
suggest that 
the effects of the self-energy and MT term 
on the value of $\rho_{ab}$ and power of the temperature dependence of $\rho_{ab}$ 
are ubiquitous. 

\begin{figure}[tb]
\includegraphics[width=123mm]{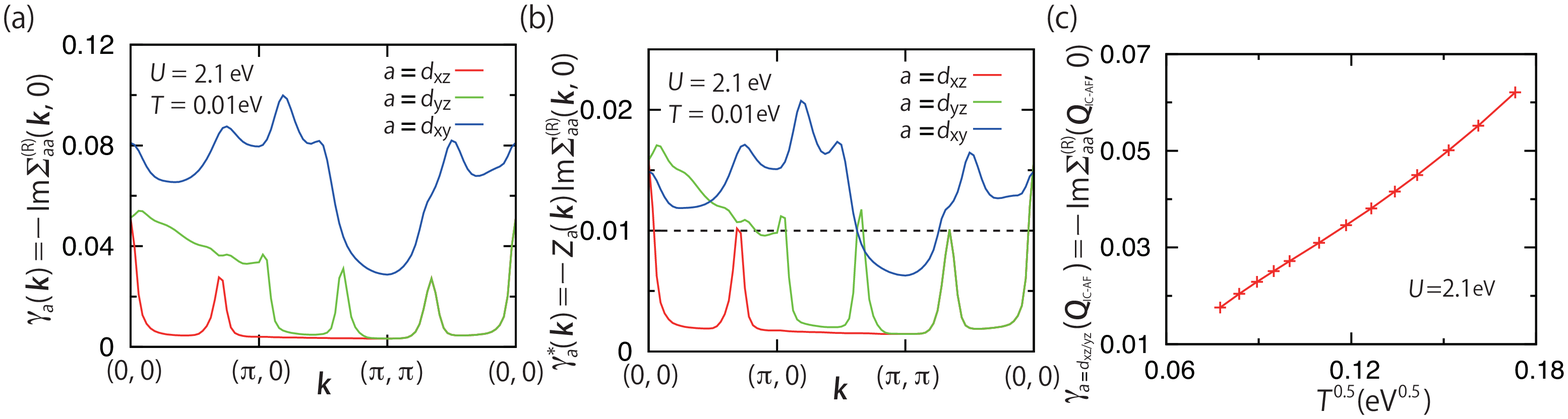}
\vspace{-12pt}
\caption{
Momentum dependences of (a) the unrenormalized QP damping and (b) the QP damping 
for each $t_{2g}$ orbital at $(U,T)=(2.1,0.01)$ (eV), 
and (c) the temperature dependence of the unrenormalized QP damping of the $d_{xz/yz}$ orbital 
at $\boldk=\boldQ_{\textrm{IC-AF}}$ against $T^{0.5}$ at $U=2.1$ eV. 
The dashed lines in panel (b) denotes $T=0.01$ eV.
}
\label{fig:Fig8}
\end{figure}

Also, 
comparing Fig. \ref{fig:Fig7} with Fig. \ref{fig:Fig4}, 
we see 
electron correlation enhances the value of $\rho_{ab}$ at each temperature. 
This is due to an increase in the unrenormalized QP dampings of the $d_{xz/yz}$ orbital 
with increasing $U$ [e.g., see Figs. \ref{fig:Fig6}(a) and \ref{fig:Fig8}(a)]. 

\begin{figure}[tb]
\vspace{-10pt}
\begin{center}
\includegraphics[width=123mm]{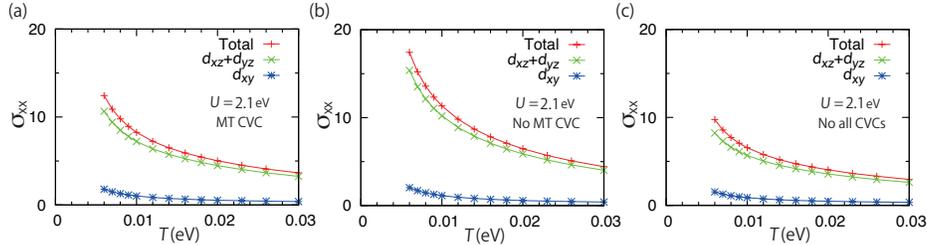}
\end{center}
\vspace{-18pt}
\caption{
Temperature dependences of the total and orbital-decomposed components of $\sigma_{xx}$ 
at $U=2.1$ eV in (a) the MT CVC case, (b) the No MT CVC case, and (c) the No all CVCs case.  
The definition of each orbital-decomposed $\sigma_{xx}$ 
is described in the main text in Sect. 3.1.  
}
\label{fig:Fig9}
\end{figure}

Then, 
in a similar way to that used at $U=1.8$ eV, 
I analyze the role of each $t_{2g}$ orbital at $U=2.1$ eV. 
Figures \ref{fig:Fig9}(a){--}\ref{fig:Fig9}(c) show 
the temperature dependences of 
the total and orbital-decomposed components of $\sigma_{xx}$ at $U=2.1$ eV 
in the three cases. 
The orbital-dependent transport holds even near the AF QCP: 
the conductions of the $d_{xz/yz}$ orbital mainly contribute to the inplane transport. 
Its mechanism is the same as that at $U=1.8$ eV, 
i.e. the smaller unrenormalized QP dampings of the $d_{xz/yz}$ orbital 
than those of the $d_{xy}$ orbital [see Fig. \ref{fig:Fig8}(a)]. 
Thus, 
this orbital-dependent transport is characteristic in some quasi-$2$D PM ruthenates. 

Moreover, 
we find from Figs. \ref{fig:Fig8}(b) and \ref{fig:Fig8}(c), respectively, 
that 
the QP damping of the $d_{xz/yz}$ orbital at $\boldk=\boldQ_{\textrm{IC-AF}}$ 
becomes the hot spot, 
and that 
the temperature dependence of 
the unrenormalized QP damping of the $d_{xz/yz}$ orbital at $\boldk=\boldQ_{\textrm{IC-AF}}$ 
is roughly proportional to $T^{0.5}$. 
Since such $T^{0.5}$ dependence near the hot spot 
causes the $T$-linear dependence of the average of the unrenormalized QP dampings 
for states along the FS\cite{Hlubina-Rice}, 
the origin of the $T$-linear $\rho_{ab}$ near the AF QCP 
is the hot-spot structure of the QP dampings of the $d_{xz/yz}$ orbital 
at momenta connected each other by the characteristic spin fluctuations of this AF QCP. 

\section{Conclusions}
In summary, 
I reviewed many-body effects\cite{NA} of $\rho_{ab}$ of 
the quasi-$2$D PM ruthenates away from and near the AF QCP 
in the FLEX approximation 
with the CVCs arising from the self-energy and MT term 
or with the CVC arising from the self-energy 
or without all the CVCs. 

The temperature dependence of $\rho_{ab}$ away from and near the AF QCP 
qualitatively agree with experiments of Sr$_{2}$RuO$_{4}$\cite{resistivity-x2} 
and Sr$_{2}$Ru$_{0.075}$Ti$_{0.025}$O$_{4}$\cite{Ti214-nFL1,Ti214-nFL2}, respectively: 
in case away from the AF QCP, 
the crossover between the $T$-linear and the $T^{2}$ dependence at about $T=0.008$ eV 
and the $T^{2}$ dependence at low temperatures are obtained; 
in case near the AF QCP, 
the $T$-linear dependence even at low temperatures is obtained. 
Here 
the main effect of the Ti substitution 
on the temperature dependence of $\rho_{ab}$ is assumed to be pushing the system 
nearer the AF QCP than Sr$_{2}$RuO$_{4}$. 

The obtained results reveal some important aspects of many-body effects 
on the resistivity of correlated electron systems. 
First, 
the overestimation of the value of $\rho_{ab}$ in the relaxation-time approximation 
and the back-flow-like effect of the MT CVC on the value of $\rho_{ab}$ 
are ubiquitous. 
It is also ubiquitous that 
the power of the temperature dependence of the resistivity 
is determined almost by the temperature dependence of the momentum- and orbital-dependent 
unrenormalized QP damping.  
The similar results have been obtained in a single-orbital Hubbard model 
on a square lattice\cite{Kon-CVC,Yanase-CVC}.  
Moreover, 
the $T$-linear resistivity near the AF QCP 
is similar to that obtained in those previous studies\cite{Kon-CVC,Hlubina-Rice,Yanase-CVC}. 
However, 
I emphasize that 
the criticality of the resistivity, the power of its temperature dependence, 
does not always connect with the criticality of spin fluctuation enhanced near a magnetic QCP 
in multiorbital systems, 
while these are always the same in single-orbital systems. 
This characteristic property comes from the facts that 
the orbital whose unrenormalized QP damping is small mainly contributes to the resistivity, 
and that 
the main orbital of the characteristic spin fluctuation of a magnetic QCP 
has the large unrenormalized QP damping. 
Since the momentum, the temperature, and the orbital dependence of 
the damping of a QP (i.e., the unrenormalized QP damping or the QP damping) 
are overlooked in Landau's FL theory 
and such momentum dependence is overlooked 
in the DMFT\cite{Georges-review,Kotliar-review}, 
the obtained results highlight 
their importance in discussing the resistivity of correlated electron systems.  

I close this paper with several remarks about the remaining issues. 
First, 
it is necessary to study the transport properties 
of other ruthenates\cite{CSRO-nFL1,Ru327-1,Ru327-2,Ru113-2} using the method I used 
and discuss the similarities and differences. 
In particular, 
the study for a quasi-$2$D ruthenate\cite{CSRO-nFL1} near a ferromagnetic QCP 
is highly desirable to understand 
the similarities and differences between 
many-body effects and role of each $t_{2g}$ orbital 
near the AF and the ferromagnetic QCP. 
Furthermore, 
it is important to analyze the transport properties of $3$D ruthenates\cite{Ru113-2} 
since comparison of the results in quasi-$2$D and $3$D ruthenates 
leads to a deep understanding of the dimensionality. 
Other remaining issues are 
the applications to other correlated electron systems 
such as transition-metal oxides\cite{MIT-review}, 
organic conductors\cite{organic-review}, 
CeCoIn$_{5}$\cite{Ce115}, and UPt$_{3}$\cite{UPt3} 
since these studies are important to deduce 
ubiquitous properties of correlated electron systems 
and characteristic properties of multiorbital systems. 
Then, 
it is intriguing to study the transport properties in a superconducting phase 
by extending the present method\cite{NA} in a PM phase 
since in some cases\cite{AL} 
the CVCs arising from not only spin fluctuations but also superconducting fluctuations 
play important roles. 
Furthermore, 
another remaining issue is to clarify the role of each $t_{2g}$ orbital 
in the thermal transport\cite{Izawa-Ru214} in the superconducting phase of Sr$_{2}$RuO$_{4}$ 
on the basis of the method where 
the orbital dependence of the damping of a QP is satisfactorily considered. 
This is because the combination of 
my result\cite{NA} and several previous studies\cite{Nomura,RG} 
suggest the existence of the difference between the main orbitals 
of the inplane transport and the superconductivity; 
if this is correct, 
we should pay attention to the effects of the orbital-dependent damping of a QP 
on the thermal transport\cite{Izawa-Ru214} 
for correct understanding of the experimental results. 

\section*{Acknowledgments}
I thank Dr. Phua Kok Khoo and Dr. Sun Han for giving me 
the opportunity to review my recent study\cite{NA} and 
waiting for the submission of this article with patience.  
All the numerical calculations were performed 
by using the facilities of 
the Supercomputer Center, the Institute for Solid State Physics,  
the University of Tokyo.

\end{document}